\documentclass[12pt,reqno]{article}
\usepackage{arxiv}
\usepackage{graphicx}
\usepackage{cite}
\usepackage{amsmath}
\usepackage{amssymb}
\usepackage{subcaption}
\usepackage{hyperref}
\usepackage{placeins}
\graphicspath{{{./}}}

\newcommand{\fref}[1]{Fig.\,\ref{#1}}
\newcommand{\tref}[1]{Table\,\ref{#1}}
\newcommand{\eref}[1]{Eq.\,(\ref{#1})}

\newcommand{\sref}[1]{Sec.\!~\ref{#1}}

\newcommand{\cref}[1]{Ref.\,\cite{#1}}

\newcommand{\ie}{{\it i.e.}\!\, }
\newcommand{\eg}{{\it e.g.}\!\, }

\newcommand{\etal}{{\it et al.} }

\newcommand{\Exp}{\mathbb{E}}
\newcommand{\Var}{\mathbb{V}}
\newcommand{\Cor}{\mathbb{C}}

\newcommand{\vb}{\mathbf{v}}
\newcommand{\kb}{\mathbf{k}}
\newcommand{\xb}{\mathbf{x}}

\renewcommand{\sb}{\mathbf{s}}
\newcommand{\yb}{\mathbf{y}}

\newcommand{\Fc}{\mathcal{F}}

\newcommand{\Lc}{\mathcal{L}}
\newcommand{\Nc}{\mathcal{N}}

\newcommand{\Cs}{\mathsf{C}}

\newcommand{\Xs}{\mathsf{X}}

\newcommand{\Ls}{\mathsf{L}}

\newcommand{\mub}{{\boldsymbol{\mu}}}

\newcommand{\epsilonb}{{\boldsymbol{\epsilon}}}
\newcommand{\varepsilonb}{{\boldsymbol{\varepsilon}}}

\newcommand{\Sigmab}{{\boldsymbol{\Sigma}}}
\newcommand{\sigmab}{{\boldsymbol{\sigma}}}

\newcommand{\dev}{{\operatorname{dev}}}

\newcommand{\NN}{\mathsf{N}\!\mathsf{N}}

\usepackage[T1]{fontenc}

\newcommand{\stress}{\sigma}
\newcommand{\strain}{\epsilon}
\newcommand{\volfrac}{\upsilon}
\newcommand{\volumefraction}{\upsilon}
\newcommand{\latent}{{{\ell}}}
\newcommand{\latentvector}{{\boldsymbol{\ell}}}

\newcommand{\phase}{\phi}
\newcommand{\propertyvector}{{\text{\bf \th}}}
\newcommand{\property      }{{\text{    \th}}}
\newcommand{\correlationkernel}{\kappa}
\newcommand{\semanticvector}{\boldsymbol{\zeta}}

\title{\bf Multiscale simulation of spatially correlated microstructure via a latent space representation}
\author{
Reese E. Jones \thanks{corresponding: \tt rjones@sandia.gov} \\
Sandia National Laboratories,\\
Livermore, CA USA
\And
Craig M. Hamel \\
Sandia National Laboratories, \\
Albuquerque, NM USA
\And
Dan Bolintineanu \\
Sandia National Laboratories, \\
Albuquerque, NM USA
\And
Kyle Johnson \\
Sandia National Laboratories, \\
Albuquerque, NM USA
\And
Robert Buarque de Macedo \\
Sandia National Laboratories, \\
Albuquerque, NM USA
\And
Jan Fuhg \\
The University of Texas at Austin, \\
Austin, TX USA
\And
Nikolaos Bouklas \\
Cornell University, \\
Ithaca, NY USA
\And
Sharlotte Kramer, \\
Sandia National Laboratories, \\
Albuquerque, NM USA
}

\begin{document}
\date{}

\maketitle{}

\begin{abstract}
When deformation gradients act on the scale of the microstructure of a part due to geometry and loading, spatial correlations and finite-size effects in simulation cells cannot be neglected.
We propose a multiscale method that accounts for these effects using a variational autoencoder to encode the structure-property map of the stochastic volume elements making up the statistical description of the part.
In this paradigm the autoencoder can be used to directly encode the microstructure or, alternatively,  its latent space can be sampled to provide likely realizations.
We demonstrate the method on three examples using the common additively manufactured material AlSi$_{10}$Mg in: (a) a comparison with direct numerical simulation of the part microstructure, (b) a push forward of microstructural uncertainty to performance quantities of interest, and (c) a simulation of functional gradation of a part with stochastic microstructure.

\end{abstract}

\section{Introduction}\label{sec:intro}

Materials with strongly heterogeneous microstructure are becoming more common with the rise of additively manufactured (AM) metals \cite{kok2018anisotropy} and the advent of exotic materials such as metamaterials \cite{jiao2023mechanical} and microarchitectured materials \cite{fleck2010micro}.
When the scales of stress and deformation are comparable to that of the microstructure through the interplay of loading and gross geometry, common simplifying assumptions such as \emph{scale separation} of the deformation field and the underlying microstructure fail to hold.
Hence many multiscale simulation methods \cite{schroder2014numerical,feyel2003multilevel} that account for microstructure but rely on scale separation become ineffective.
Likewise, classical micromechanics theory is mostly concerned with the limits of averages/mean properties \cite{nemat2013micromechanics}, although finite-size effects and spatial correlations need to be accounted for.
This poses a challenge for the efficient design of structures that accounts for microstructure \cite{fullwood2010microstructure} since direct numerical simulation (DNS) of the microstructure of an entire part is typically infeasible.
Furthermore, only the spatial statistics of the microstructure are usually known, not the exact microstructural configuration, which requires a design process based on sampling multiple realizations of the microstructure to account for this uncertainty.

To address this challenge we propose a multiscale simulation method that incorporates the spatial correlations of the underlying microstructure and, thereby, some of the finite-size effects that complicate homogenization in this scenario.
The goal is to enable the design of components with interacting gross and microstructural scales under microstructural uncertainty by accelerating the forward propagation of this uncertainty.
The method uses a variational autoencoder (VAE)-like neural network (NN) \cite{wang2020deep} to encode microstructure into a compact latent vector and thereby to effective properties of the microstructural sample.
These properties, in turn, determine the response of finite elements representing the neighboring, correlated high-dimensional microstructural samples.
In effect the NN provides the structure-property relationship for the material and, through an intermediate structure-to-latent encoding, a compact representation of the microstructural features relevant to property prediction.
By sampling the latent distribution of statistically representative samples, the method can be used generatively for uncertainty quantification (UQ); alternatively, if the full microstructure of a component is available, the VAE can be used to encode it into fields of corresponding latent descriptors and properties.
Our use of interpretable properties is expedient since it allows the incorporation of the method into large-scale simulators with minimal modification.

In the next section, we provide background to give context for the proposed method.
Then, in \sref{sec:problem}, we provide the problem setting.
\sref{sec:structure} describes the methods we use to create statistically consistent realizations.
In \sref{sec:data} we give details of the training data generation, and in \sref{sec:architecture} we describe the NN architecture that provides the engine for the proposed multiscale method.
Then in \sref{sec:demonstrations} we provide three demonstrations motivated by
laser-powder-bed-fusion additive manufacturing (L-PBF AM) with AlSi$_{10}$Mg \cite{specht2021shock,takata2018size}:
(a) a comparison of the multiscale response to that of a corresponding DNS,
(b) forward propagation of microstructural uncertainty for a case where deformation gradients interact directly with microstructure, and
(c) an illustration of the use of the method to simulate random materials with functional gradation.
Lastly, \sref{sec:conclusion} provides a summary of conclusions and avenues for future work.

\section{Related work}\label{sec:background}

This work builds upon developments in microstructural characterization, synthetic realization generation, homogenization, and related fields to deliver an NN-powered multiscale simulation method that embeds correlated microstructural features.
Each of these fields has a long history of development.
Here we give a brief topical review of developments relevant to the present work.

Classical microstructural characterization can be seen as a form of encoding high-dimensional structures into a few interpretable statistics.
For instance, Kalidindi and coworkers have been particularly productive in the use of 2-point statistics to describe the spatial correlations of experimentally observed microstructures \cite{niezgoda2010optimized,kalidindi2011microstructure,niezgoda2013novel,kalidindi2015hierarchical,cecen2016versatile}.
These spatial statistics provide the basis for numerous structure generation techniques such as Gaussian random fields (GRFs) \cite{williams2006gaussian,gabrielli2006statistical,kroese2014spatial} and Karhunen-Lo\`eve expansions (KLEs) \cite{phoon2005simulation,allaix2013karhunen,cho2013karhunen,panunzio2018large,daw2022overview} through spectral connections via the Wiener-Khinchin and related theorems.
General latent encoding of high dimensional inputs is also a common task in machine learning.
The Variational auto-encoder (VAE) \cite{kingma2013auto,doersch2016tutorial,kingma2019introduction} has seen wide-spread use in this task.
As an autoencoder, the VAE reduces a high dimensional input, such as an image, into a compact latent space, and then decodes this latent encoding to reproduce the input using an hourglass-like stack of NN layers.
A VAE is variational in the sense that parameters of approximate latent distribution are optimized to match the training data.
Typically a multivariate Gaussian is used to represent the latent distribution but others, such as the so-called \emph{concrete} distribution \cite{maddison2016concrete}, may be more appropriate for encoding binary images/structures \cite{mena2019binary,rolfe2016discrete}.
Chen and co-workers \cite{wang2020deep} invented a property-VAE (pVAE) which augments a standard VAE with a regressor devoted to the secondary task of predicting the properties of the input structure.
They showed that the VAE has some facility in organizing the latent space so that there is a smooth gradient of properties.
The secondary task also emphasized the fact that the VAE is seeking a multi-objective solution \cite{higgins2017beta}  in terms of input reconstruction loss, the divergence between the surrogate latent distribution and the data distribution, and (c) the regression error of the predicted properties, which need to be balanced in a single objective.

The homogenization of the properties and response of heterogeneous random media is a mature field of research \cite{nemat2013micromechanics,mura2013micromechanics,torquato2013random} and provides the foundation for multiscale simulation.
The Hill-Mandel \cite{hill1963elastic,mandel1972plasticite} condition, postulating the equivalence of power expended at the micro and macro scale plays a central role in many multiscale methods.
Although generally concerned with ensemble averages and large-sample limits, the finite-size effects/variability inherent in small samples \cite{ranganathan2008scaling,ostoja2007microstructural} as well as the correlation structures \cite{ostoja1999microstructural} are also themes in this discipline.
Computational multiscale methods  \cite{kevrekidis2003equation,feyel2003multilevel,weinan2007heterogeneous} typically ignore these effects and appeal to scale separation of the microstructure and the other length scales in the problem of interest.
In fact, the well-cited FE$^2$ method \cite{feyel2003multilevel,schroder2014numerical} is predicated on scales being well separated.
However, recently Chen, Guilleminot,  and Soize \cite{chen2024concurrent} published a treatment of random elastic media where spatial coherence of the realizations was incorporated.
The earlier work of Girolami \etal \cite{girolami2021statistical} also provided a probabilistic learning framework applied to material modeling.
A number of recent reviews by Peng \etal  \cite{peng2021multiscale}, Fish \etal \cite{fish2021mesoscopic}, and Bishara \etal \cite{bishara2023state} provide a more comprehensive view of the application of machine learning to the multiscale problem.

The treatment of the material uncertainty due to microstructure variability \cite{rizzi2019bayesian,khalil2021modeling} and design under this uncertainty \cite{bessa2017framework} are closely related tasks to multiscale modeling \cite{rizzi2019bayesian,khalil2021modeling}, and they have recently seen the rise of purely data-driven approaches \cite{kirchdoerfer2016data,eggersmann2019model,carrara2020data,karapiperis2021data},
As a counterpoint, machine-learning discovery of models of material response \cite{jones2022neural,flaschel2022discovering,flaschel2023automated,liu2023learning} can provide data reduction of a complex material response as a function of microstructure \cite{frankel2019predicting,frankel2020prediction,he2021deep}.
The data reduction of the response of microstructural samples can be taken to the point of obtaining material properties from the samples (aka parameters of chosen models) \cite{heidenreich2023modeling}.
Each of these approaches involving encoding the structure-response map have clear advantages and disadvantages in terms of generality, computational expense and ease of implementation.

\section{Problem} \label{sec:problem}

In pursuit of an efficient, acceptably accurate solution to a boundary value problem with pronounced heterogeneity, concurrent multiscale (MS) methods typically condition a field of small-scale problems on a coarse kinematic field and use these responses to drive the solution of the overall problem.
Although there may be many (ill-defined) scales of heterogeneity, the scale-separation assumption of traditional multiscale methods allows these small-scale boundary value problems to represent independent samples of the microstructure.
In this work, we provide an approximation for the case where the range of influence of the scales are overlapping by accounting for the spatial correlation of the underlying structure.
This case is prevalent when there is interplay between the deformation-gradient field and characteristic structural scales, as when the loading and/or the nominal geometry produce deformation gradients that change over a scale comparable to that of the microstructure \cite{teichert2022sensitivity}.
In these scenarios, traditional MS will fail to be predictive, while the proposed approach should bring the multiscale solution closer to that of the DNS treatment that resolves the microstructure.

\begin{figure}[htb!]
\centering
\includegraphics[width=0.45\textwidth]{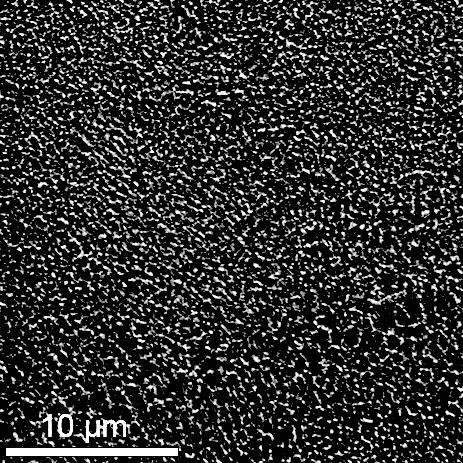}
\caption{AlSi$_{10}$Mg microstructure
black: Al, white: Si
courtesy of Sara Dickens and Paul Specht, Sandia National Laboratories \cite{specht2021shock}.
}
\label{fig:AlSi10Mg_microstructure}
\end{figure}

Since microstructure cannot be controlled precisely, it is typically characterized and modeled statistically.
Although many microstructures appear random at some scales, such as the micrograph of  AlSi$_{10}$Mg shown in \fref{fig:AlSi10Mg_microstructure}, they typically have a spatial correlation that can be measured with 2-point statistics \cite{kalidindi2015hierarchical}.
These spatial statistics are a form of latent encoding that we will exploit in synthetic structure generation described in \sref{sec:structure}.

We will assume that a well-defined set of properties $\propertyvector$ exists at the smallest relevant structural scale.
This is plausible for observable structural properties, such as phase concentration, while mechanical properties are harder to probe directly at small scales but are clearly dependent on these observables.
For phase-separated mixtures and alloys, simple mixture rules give an expectation of derived quantities.
If the contributions are independent, such as for mass density, these rules are exact.
For other, more complex, deformation-dependent quantities we need a model to predict the properties that control the response of the microstructure.

\begin{figure}[htb!]
\centering
\includegraphics[width=0.45\textwidth]{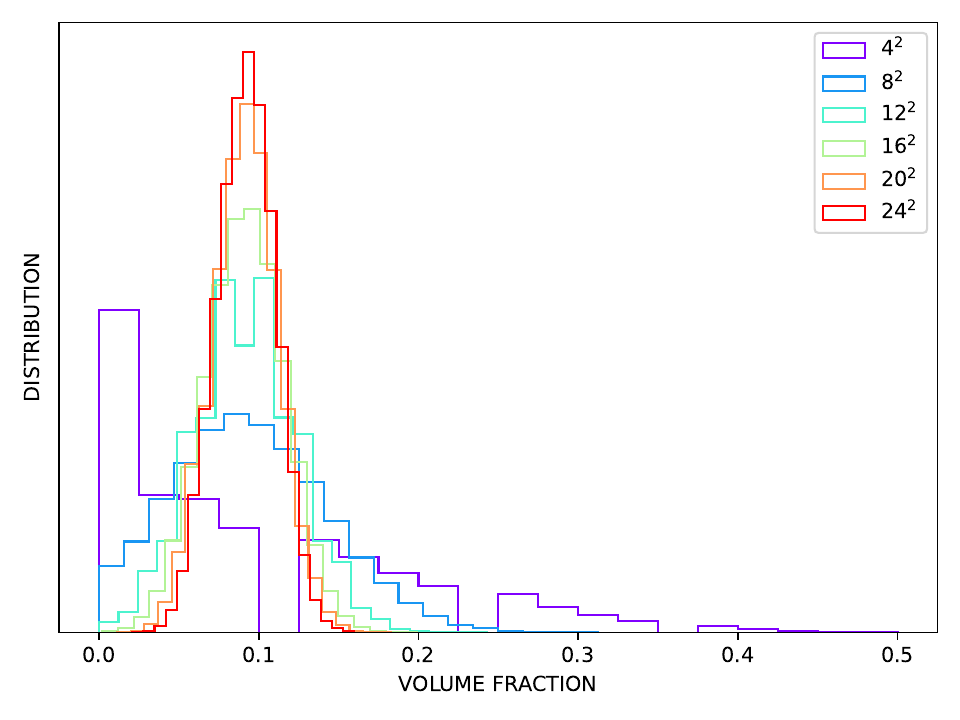}
\includegraphics[width=0.45\textwidth]{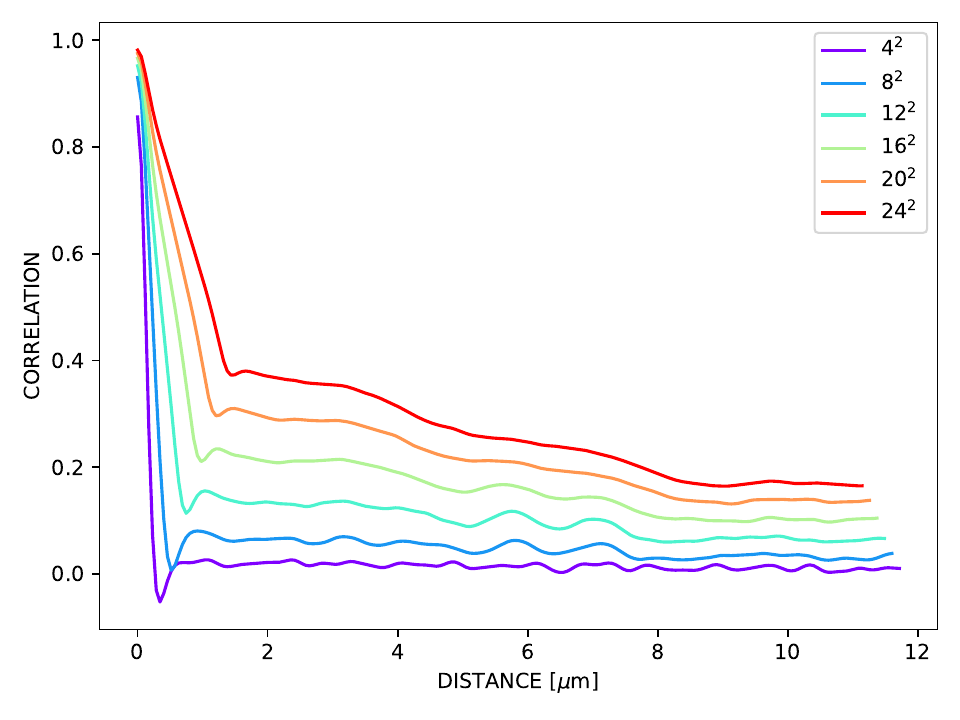}
\caption{Volume fraction distribution and correlation of AlSi$_{10}$Mg microstructure
}
\label{fig:AlSi10Mg_stats}
\end{figure}

Unlike in an homogeneous material, for a body $\Omega$ with microstructure, the property $\property$
\begin{equation} \label{eq:sve_property}
\property_V \equiv \frac{1}{V} \int_{\square} \property(\xb) \, \mathrm{d}V
\end{equation}
of a finite-sized sample $\square$ with volume $V$ can vary by location.
A \emph{stochastic volume element} (SVE) is the common term for a microstructural sample that is smaller than the limit
\begin{equation} \label{eq:prop_mean}
\bar{\property} \equiv
\Exp[\property_V] =
\lim_{V\to\infty} \property_V
\end{equation}
at which the particular property reaches its mean value $\bar{\property}$.
The smallest sample that attains this limit, to a given accuracy, is called a \emph{representative volume element} (RVE).
For SVEs, there can be significant variance in the properties:
\begin{equation} \label{eq:prop_var}
\Var[\property_V] = \frac{1}{V_\Omega} \int_\Omega (\property_V(\xb) - \bar{\property})^2 \, \mathrm{d}V  \propto \frac{1}{V} \gg 0
\end{equation}
We expect the variance in properties with sample size to follow a central limit trend, in the sense that it is roughly inversely proportional to sample size since each sample property is effectively an average of contributions throughout the sample.
\fref{fig:AlSi10Mg_stats} shows the distribution of binary phase fraction and spatial correlations for a sequence of sample sizes associated with the material shown in \fref{fig:AlSi10Mg_microstructure}.
Clearly, the distribution narrows and the correlation length increases with increasing sample size.

Furthermore, the integral over the region $\square$ in \eref{eq:sve_property} can be considered a filter of an underlying random process $\property(\xb)$ in the sense that $\property_V$ is the convolution of the field $\property(\xb)$ at the smallest scale with top hat kernel with extent given by the SVE size $\square$.
Hence for a coherent structure, such as that shown in \fref{fig:AlSi10Mg_microstructure}, the SVE properties at any location are clearly correlated with the neighboring values.
The 2-point correlation
\begin{equation} \label{eq:prop_covar}
\Cor[\property(\xb)]
= \Exp[\property_V(\mathbf{0}) \property_V(\mathbf{\xb})]
= \frac{1}{V_\Omega} \int_\Omega \property_V(\yb) \property_V(\yb+\xb)  \mathrm{d}\yb
\end{equation}
is a common statistic used to characterize this dependence.
Typically correlations decay with distance, with the notable exception of periodic structures, and hence have a characteristic length.
Also, it stands to reason that, as the SVE property estimate approaches the RVE limit, the variance of the estimate will decrease, and the correlation length will increase.
In this work, we assume $\Cor[\property(\xb)] = \Cor[\property(\|\xb\|)]$ to be isotropic for simplicity in the following developments.
Note that the variance in \eref{eq:prop_var} $\Var[\property] = \Cor[\property(\mathbf{0})]$ i.e. the expected variance of a property in a sample is given by the self-correlation.
\fref{fig:AlSi10Mg_stats} shows the spatial correlation of the average phase of SVEs of increasing size corresponding to the microstructure shown in \fref{fig:AlSi10Mg_microstructure}.
The rapidly decaying primary correlation is associated with the average feature size, and the dip in correlation is likely due to one phase being anti-correlated with the other, while the longer decaying trend indicates a larger scale order.
More generally this figure illustrates how: (a) the variance of a mean property decreases with increasing sample size and (b) the spatial correlation of this mean property increases with increasing sample size.
The limit being the RVE value of the property that is equivalent everywhere and hence has a correlation length as large as the body.

If the mean \eqref{eq:prop_mean} and covariance \eqref{eq:prop_covar} do not depend on location in the larger body $\Omega$, these statistics are sufficient to describe a stationary second-order process.
More exposition of spatial statistics can be found in \cref{gelfand2010handbook} for instance.
These two statistics are generally insufficient to determine the mechanical properties, but all aspects of the high-dimensional structural configuration space are not needed to determine the necessary properties.
In \sref{sec:architecture} we will describe how we can reduce the structure-property relationship to a compact latent space.

\section{Microstructure generation} \label{sec:structure}
Since, in applications ranging from component design to part forensics, we typically only have statistical characterization available, we need structure generation methods.
For our purposes, we need to generate synthetic realizations with the measured spatial statistics in order to push forward the variability to performance metrics.
There are a multitude of methods to generate microstructures that obey given spatial statistics.
We employ two methods, in different capacities, which generate fields with a given spatial correlation and point-wise expectation: (a) a Gaussian random field (GRF) and (b) a Karhunen-Lo\`eve expansion (KLE).
We use a GRF to generate SVEs and structures for direct numerical simulation (DNS), and we use KLE to impart spatial correlation of latent vectors on meshes used to perform multiscale simulation.
Each method relies on a correlation function that we assume, for simplicity, is isotropic.
For structure generation we employ a correlation function representation with multiple decay / correlation lengths
\begin{equation} \label{eq:corr_kernel}
\correlationkernel(\|\xb-\xb'\|)
= \sum_i w_i  \exp\left( -\frac{\| \xb-\xb'\|^p}{\ell_i^p} \right)
\end{equation}
with $\sum_i w_i = 1$ and $p>0$.

A Gaussian random field \cite{gabrielli2006statistical,kroese2014spatial} is a random field with given spatial correlation, and can be readily constructed on a structured grid using fast Fourier transforms.
In essence this involves the convolution of a noise process $\eta$ without spatial dependence with a kernel $\correlationkernel$.
\begin{equation} \label{eq:grf}
\varphi(\xb)
=   \Fc^{-1}_{\kb\to\xb} \left[\Fc_{\xb\to\kb}[\eta] \sqrt{\left|\Fc_{\xb\to\kb} [\correlationkernel]\right|} \right]
\end{equation}
where $\Fc$ is a spatial Fourier transform, and $\left|\Fc_{\xb\to\kb} [\correlationkernel]\right|$ is the spectral density of the spatial correlation.
Typically $\eta$ is Gaussian white noise, and $\correlationkernel$ is a correlation kernel resembling \eref{eq:corr_kernel}.
From the form of \eref{eq:grf} it is clear that a GRF is defined by the power spectral density of the spatial correlation, and the statistics of the point process are given by $\eta$.
With this Fourier transform construction, a GRF is particularly suited for generating periodic microstructures.
Although more sophisticated methods exist \cite{ma2009construction,dynkin1984gaussian,bocchini2008critical}, \eref{eq:corr_kernel} is sufficient for this exposition.

The Karhunen-Lo\`eve expansion (KLE) is another method that uses kernel $\correlationkernel$, which is more adaptable to meshes but also more computationally expensive than GRFs.
We will use a KLE to sample the latent space generated by the pVAE described in \sref{sec:architecture}.
A concise exposition of the KLE applied to this task will be given in \sref{sec:uq}.

Since these methods generate continuous fields, to create binary structures,
a threshold $\varepsilon$ that depends on a target volume fraction $\volumefraction$ is typically employed
\begin{equation}
\phase(\xb) = \begin{cases}
0 & \text{if} \  \varphi(\xb) > \varepsilon \\
1 & \text{else}
\end{cases}
\end{equation}
Adding multiple levels provides a direct extension to produce multiphase alloys.

\begin{figure}[htb!]
\centering
\includegraphics[width=0.95\textwidth,height=0.15\textwidth]{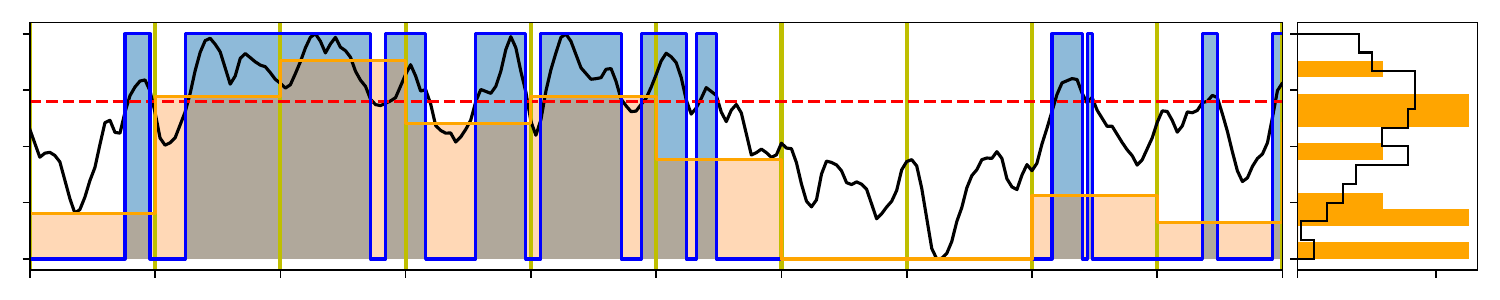}
\caption{1D illustration of the microstructure generation process.
A realization of the underlying random process is depicted by the black curve,
the binarization threshold is indicted by the horizontal dashed red line, and the resulting binary phase is shown as a blue histogram, while the volume fractions of SVEs are shown as an orange histogram.
The histograms on the right show the distribution of the random process (black) and the volume fractions (orange) for the selected SVE size.
}
\label{fig:1d_illustration}
\end{figure}

\fref{fig:1d_illustration} illustrates the overall procedure in one dimension.
The black line is the GRF realization on the 1D domain.
The red dashed line is the threshold $\varepsilon$, which results in the binary field in blue.
The orange steps indicate the volume fraction per cell, which have the distribution shown in the side panel.
The volume fraction distribution is erratic for this small sample but has the most mass at the expected value.

\begin{figure}[htb!]
\centering
\includegraphics[width=0.95\textwidth]{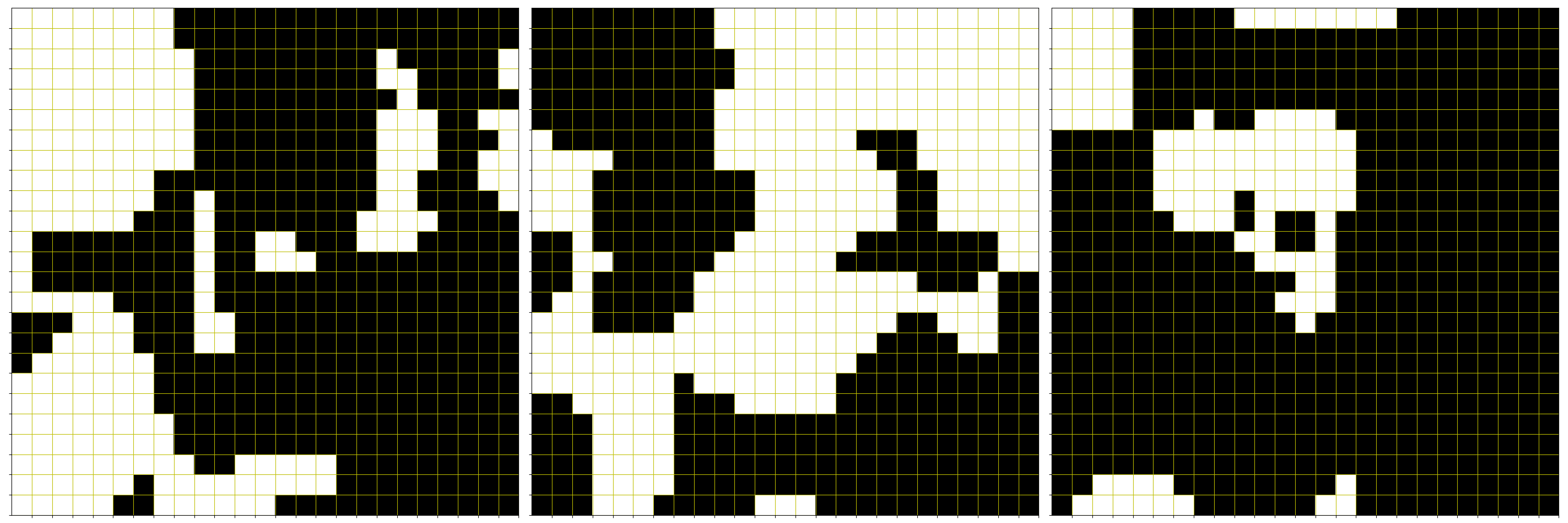}
\caption{Representative SVEs
Dividing lines are at voxel size.
}
\label{fig:sves}
\end{figure}

\begin{figure}[htb!]
\centering
\includegraphics[width=0.95\textwidth]{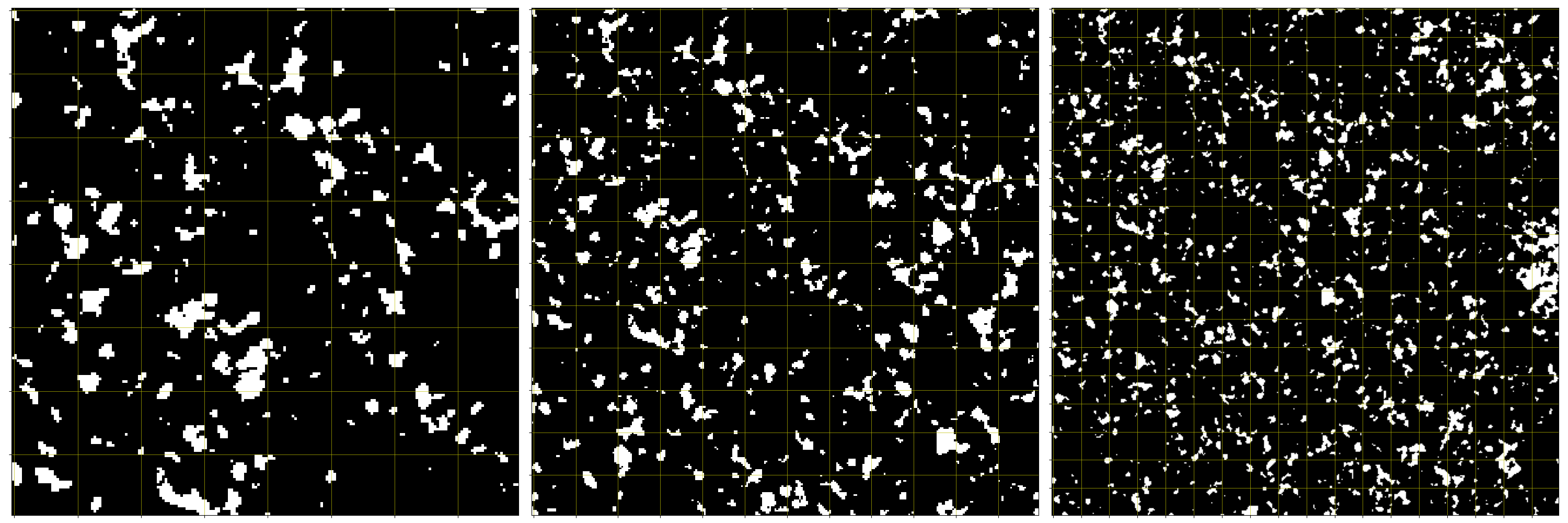}
\caption{Representative larger scale samples.
Dividing lines are at the  SVE size shown in \fref{fig:sves}.
}
\label{fig:dns}
\end{figure}

\fref{fig:sves} shows representative SVEs constructed with correlation
\begin{equation}
\correlationkernel(r) = 0.95  \exp(-r/1.5) + 0.05 \exp(-r/15.0) \ ,
\end{equation}
which resembles that of \fref{fig:AlSi10Mg_stats},
where $r$ is in units of voxels, and the mean of the process is $\bar{\varphi} = 0.1$
In both the SVE and the larger samples shown in \fref{fig:dns} coherent structures and structural scales can be observed.

\section{Dataset} \label{sec:data}

Our exemplar material is nominally AlSi$_{10}$Mg, which we simplify to a binary mixture of 10\% Si and 90\% Al.
We assume the two phases are segregated as in \fref{fig:AlSi10Mg_microstructure}.
Over the range of interest, Si is modeled as elastic, and Al is modeled as elastic-plastic with properties given in \tref{tab:dns_properties}.
Specifically, we used a standard J2 elastic-plastic model with isotropic hardening for Al.
The Al hardening follows the often-employed Swift-Voce (SV) rule \cite{voce1948relationship,swift1952plastic} :
\begin{equation} \label{eq:voce}
h(\epsilon_p) = Y + H \epsilon_p + A (1-\exp(-B \epsilon_p)),
\end{equation}
which is essentially linear hardening augmented with an exponential trend.
In \eref{eq:voce}
\begin{equation}
\epsilon_p = \sqrt{\left( \frac{2}{3} \| \dev \epsilonb_p \| \right)}
\end{equation}
is the equivalent plastic strain, and $\epsilonb_p$ is the plastic strain.

We used tensile loading to extract the relevant properties of the SVE samples and ignore the anisotropy induced by the microstructural configurations for simplicity although it is likely a source of error.
We used periodic displacement boundary conditions to reduce the potential for boundary effects overwhelming the bulk response.

The volume fraction of Si $\volfrac$ of each SVE was determined directly from the structure.
The mechanical properties were extracted from the stress-strain response using standard methods.
Young's modulus $E$ was fit to the mean slope $\stress_{11} / \strain_{11}$, and  Poisson's ratio $\nu$ to the mean lateral contractions $(\strain_{22} + \strain_{33}) / 2 \strain_{11}$.
Using the fitted $E$ the plastic regime was determined by the 0.2 \% offset in strain rule.
All the data in the plastic regime was used to fit the Voce rule \eref{eq:voce} to determine yield $Y$ and the three hardening parameters $H$,$A$,$B$ for every SVE.
The similarity of the SVE response to the SV rule enabled a high accuracy representation, although this generally may not be valid.
Note that the linear and exponential components of the SV hardening rule can confound each other; and so care in fitting is needed, \eg good starting guesses to optimizer in region of attraction or regularization/penalization of irregular fits.
Although generally not feasible, for the present case regressing to a known functional form to extract property features was an accurate reduction of the response complexity.
The range of effective properties and the mean standard errors of the fits are given in \tref{tab:sve_properties}.

\begin{figure}[htb!]
\centering
\includegraphics[width=0.60\textwidth]{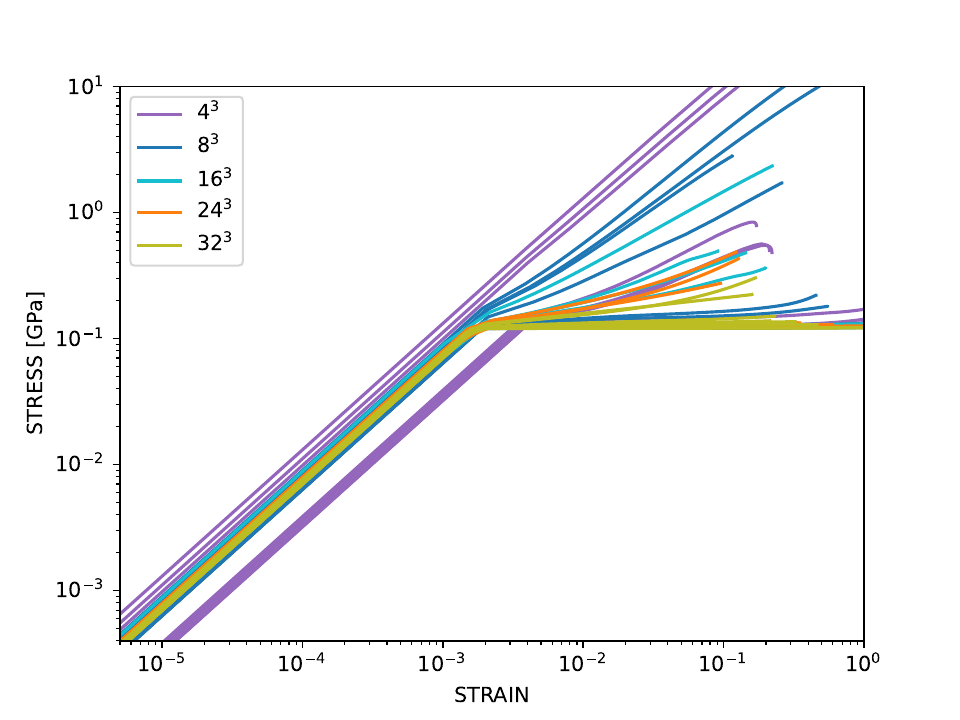}
\caption{
Stress-strain response for a sampling of different SVE sizes.
Note the log scale on both axes.
}
\label{fig:response_size}
\end{figure}

Prior to sampling a large ensemble of SVEs and their properties for a training dataset,  we explored the size dependence of the properties of SVEs with 4$^3$, 8$^3$, 16$^3$, 24$^3$, 32$^3$ elements.
These SVEs were extracted from a DNS structure at least 10 times as large as an SVE with the selected volume fraction,
\fref{fig:response_size} shows a sampling of the stress-strain response curves on a log-log scale to show the range of responses.
For the smallest SVE size, 4$^3$, there is a significant propensity to sample high Si volume fractions, and for these samples, the stress-strain response is nearly linear.
\fref{fig:response_size} also illustrates the decreasing variance in response with increasing SVE size.

\begin{figure}[htb!]
\centering
\includegraphics[width=0.32\textwidth]{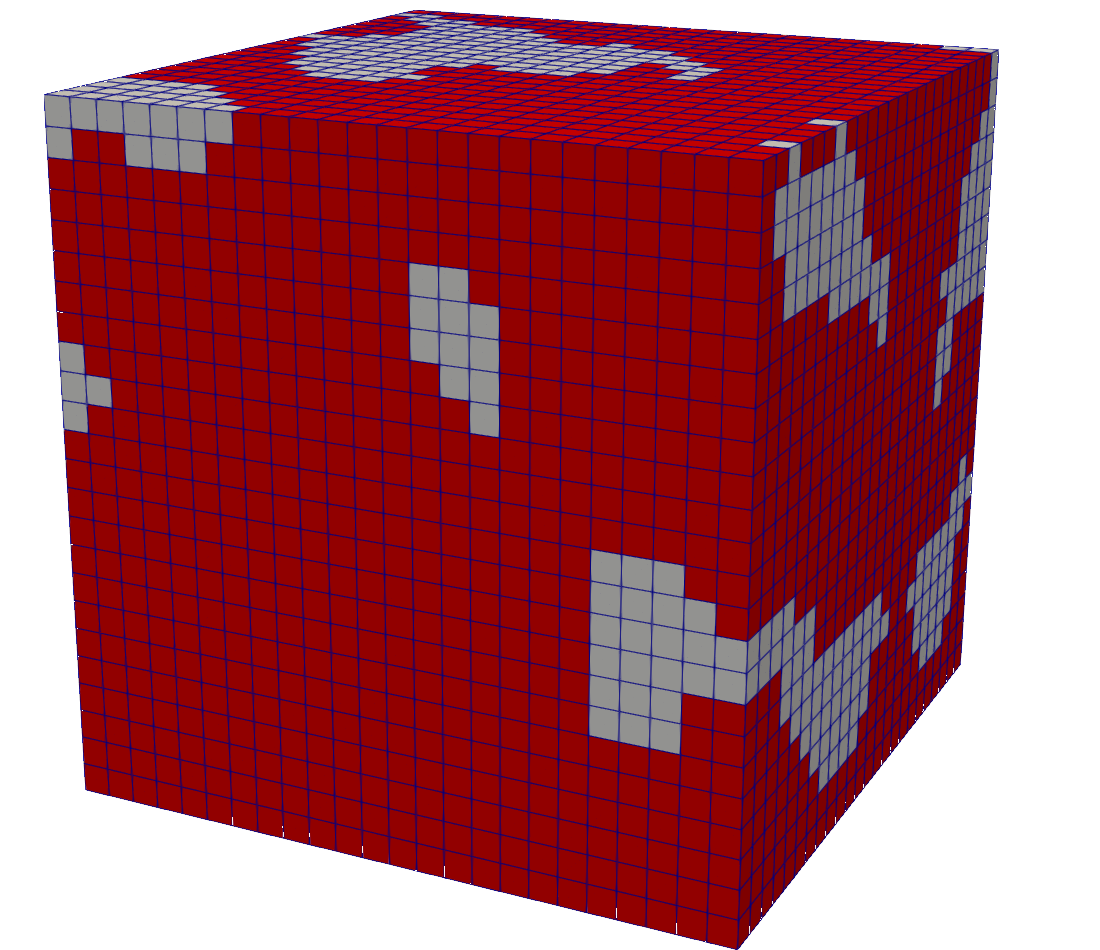}
\includegraphics[width=0.32\textwidth]{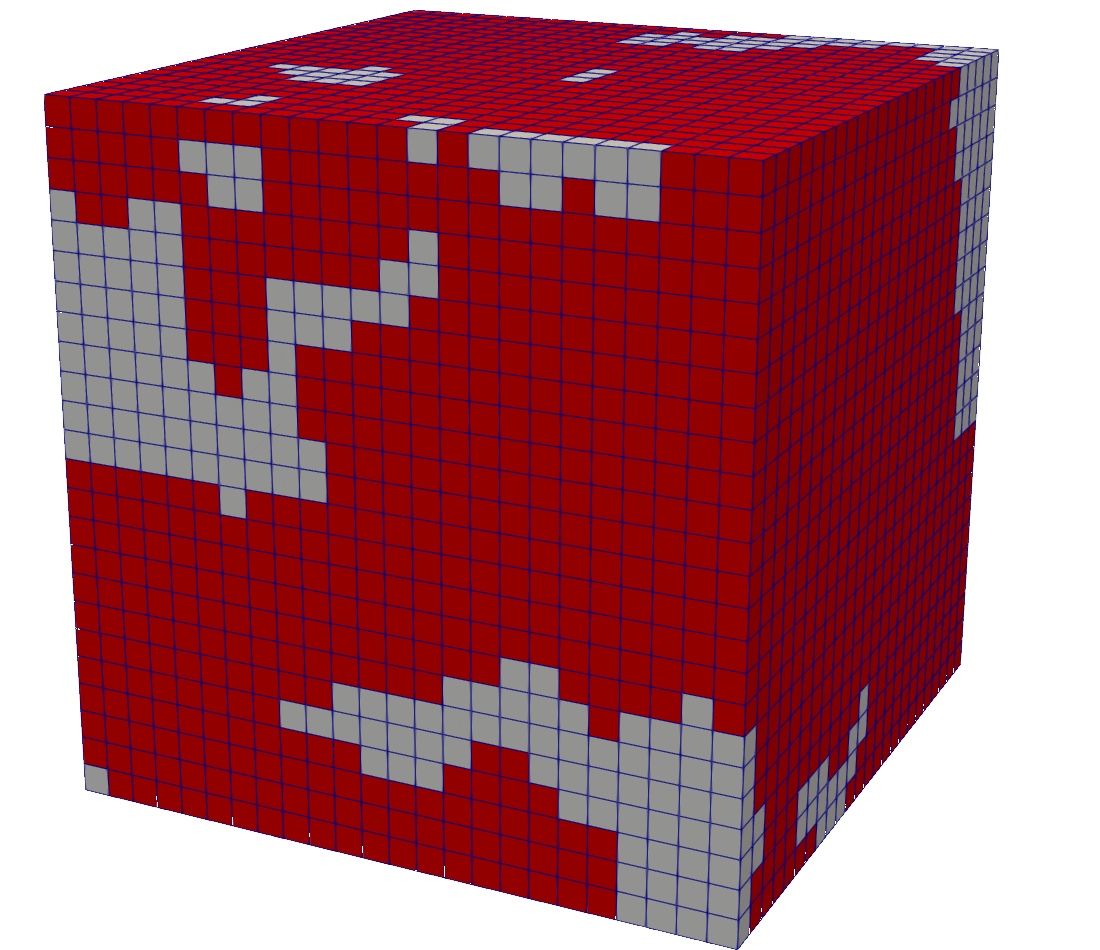}
\includegraphics[width=0.32\textwidth]{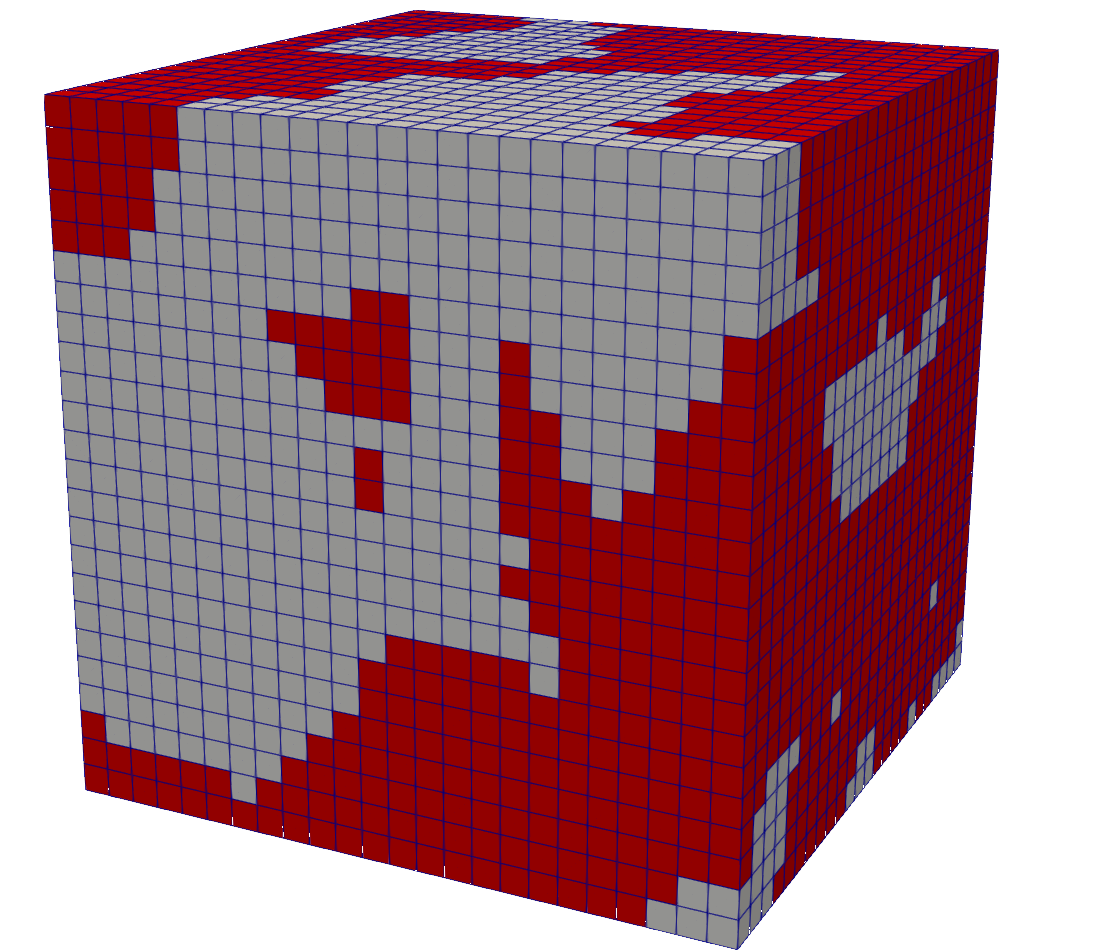}

\includegraphics[width=0.32\textwidth]{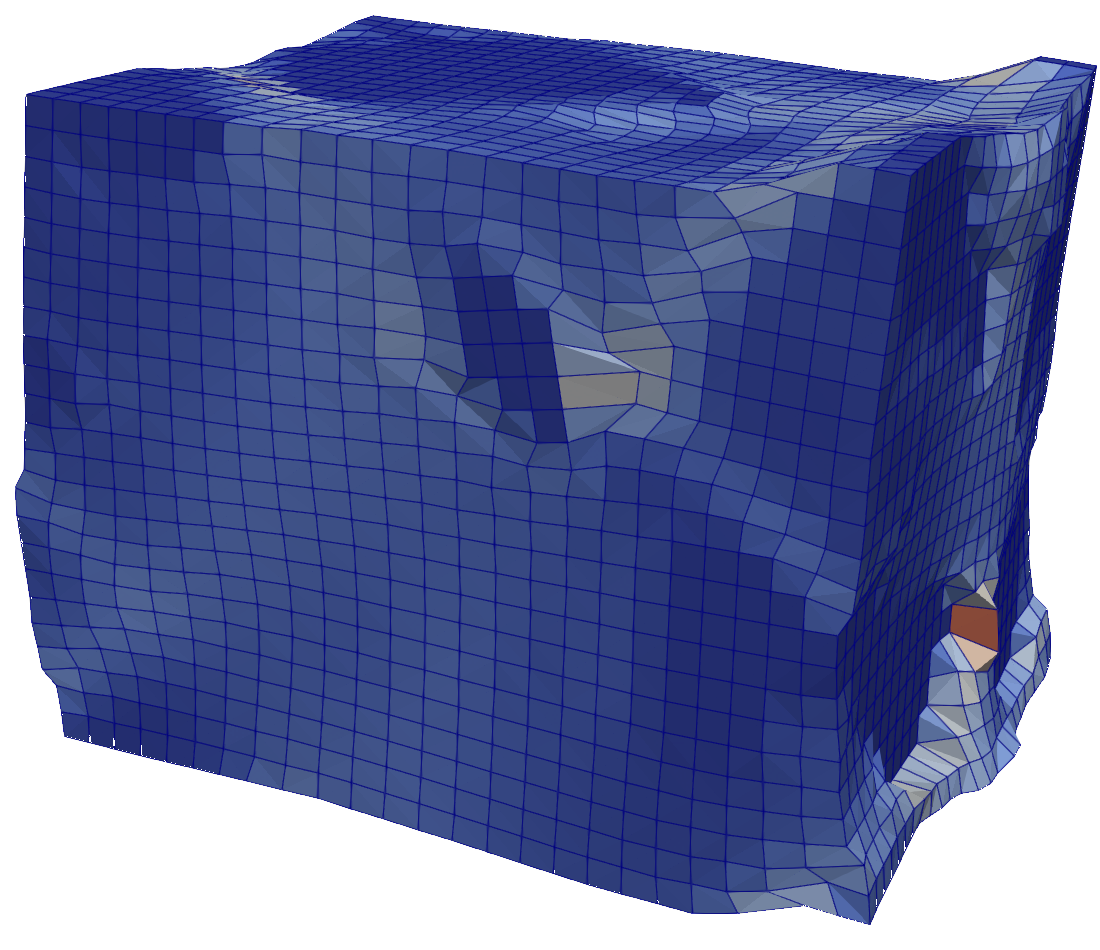}
\includegraphics[width=0.32\textwidth]{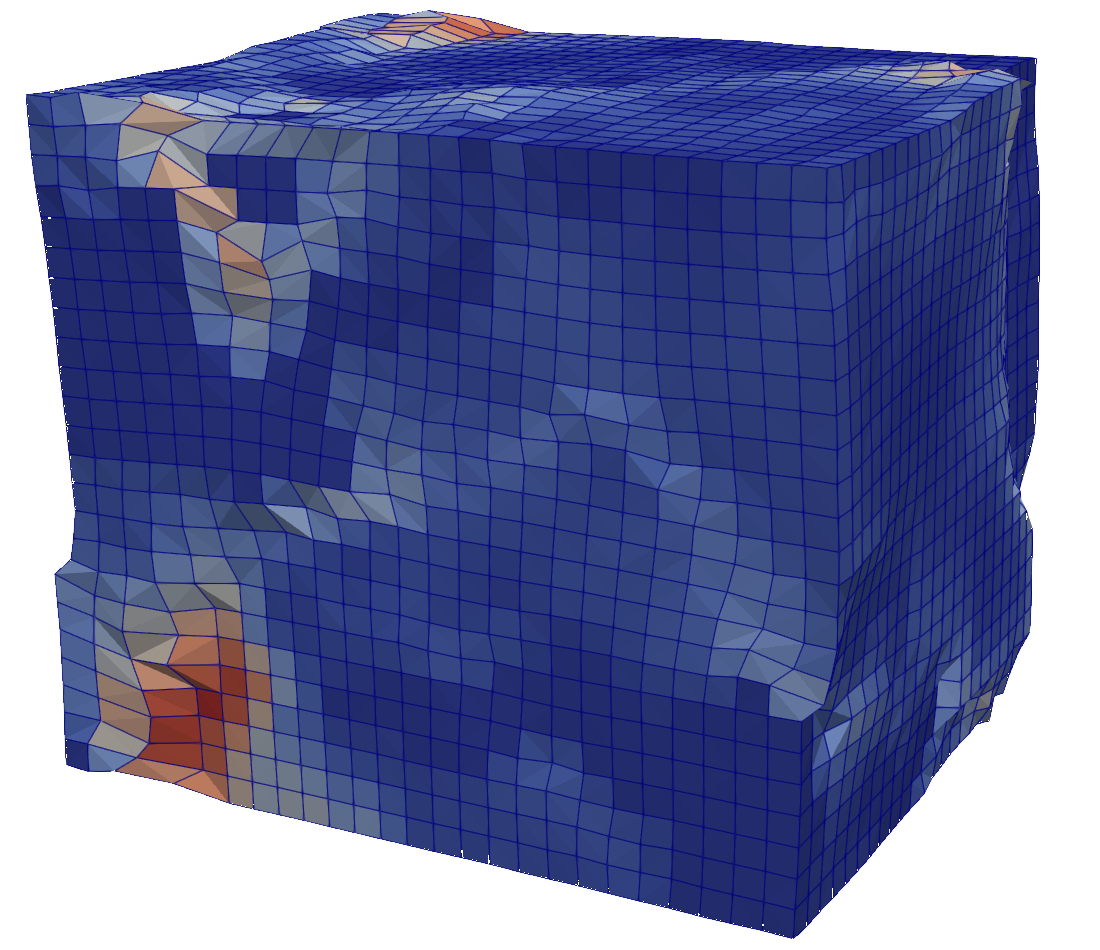}
\includegraphics[width=0.32\textwidth]{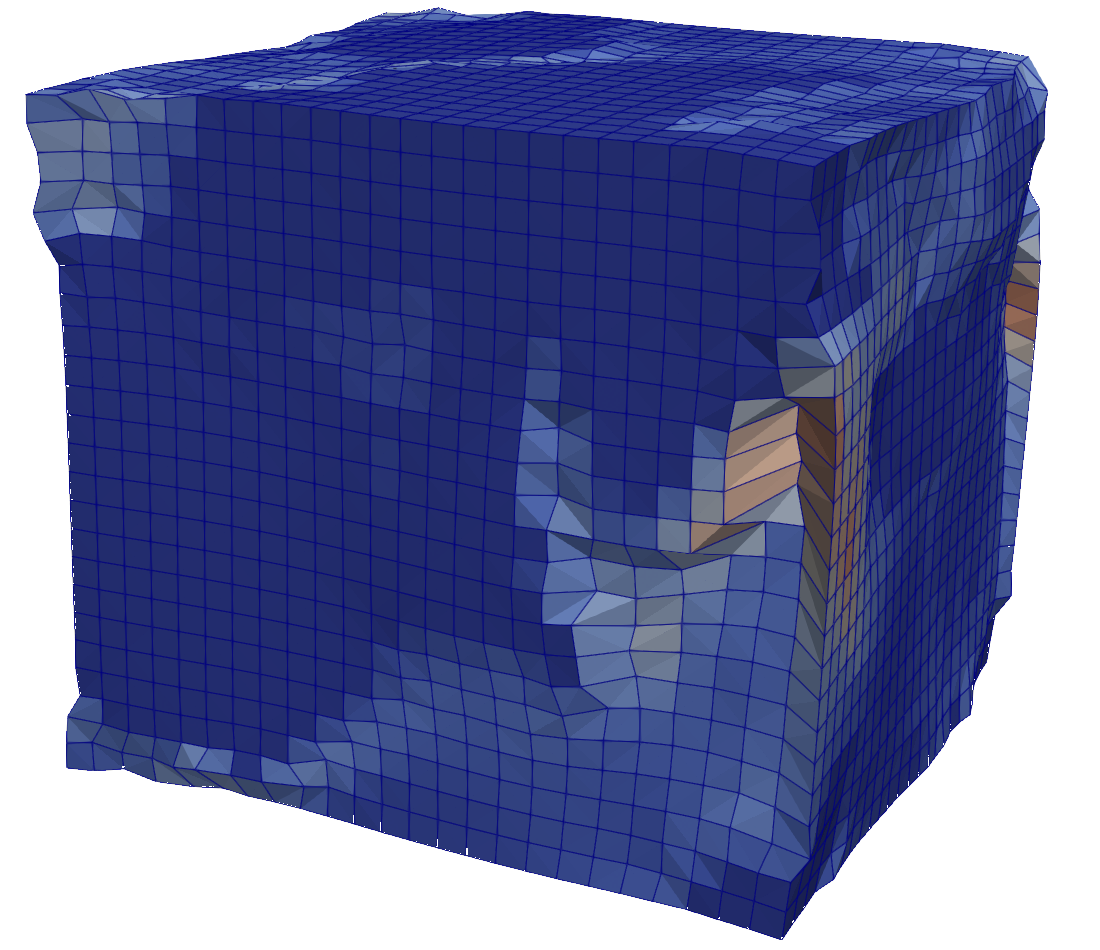}
\caption{Realizations of 24$^3$ cells for volume fractions $\volfrac$ = 0.1(left), 0.2 (middle), 0.3(right).
Phase (upper row)  red:Al, gray:Si.
Equivalent plastic strain (lower row) at failure 0.0 blue, red: 1.5
}
\label{fig:realizations}
\end{figure}

\begin{figure}[htb!]
\centering
\includegraphics[width=0.65\textwidth]{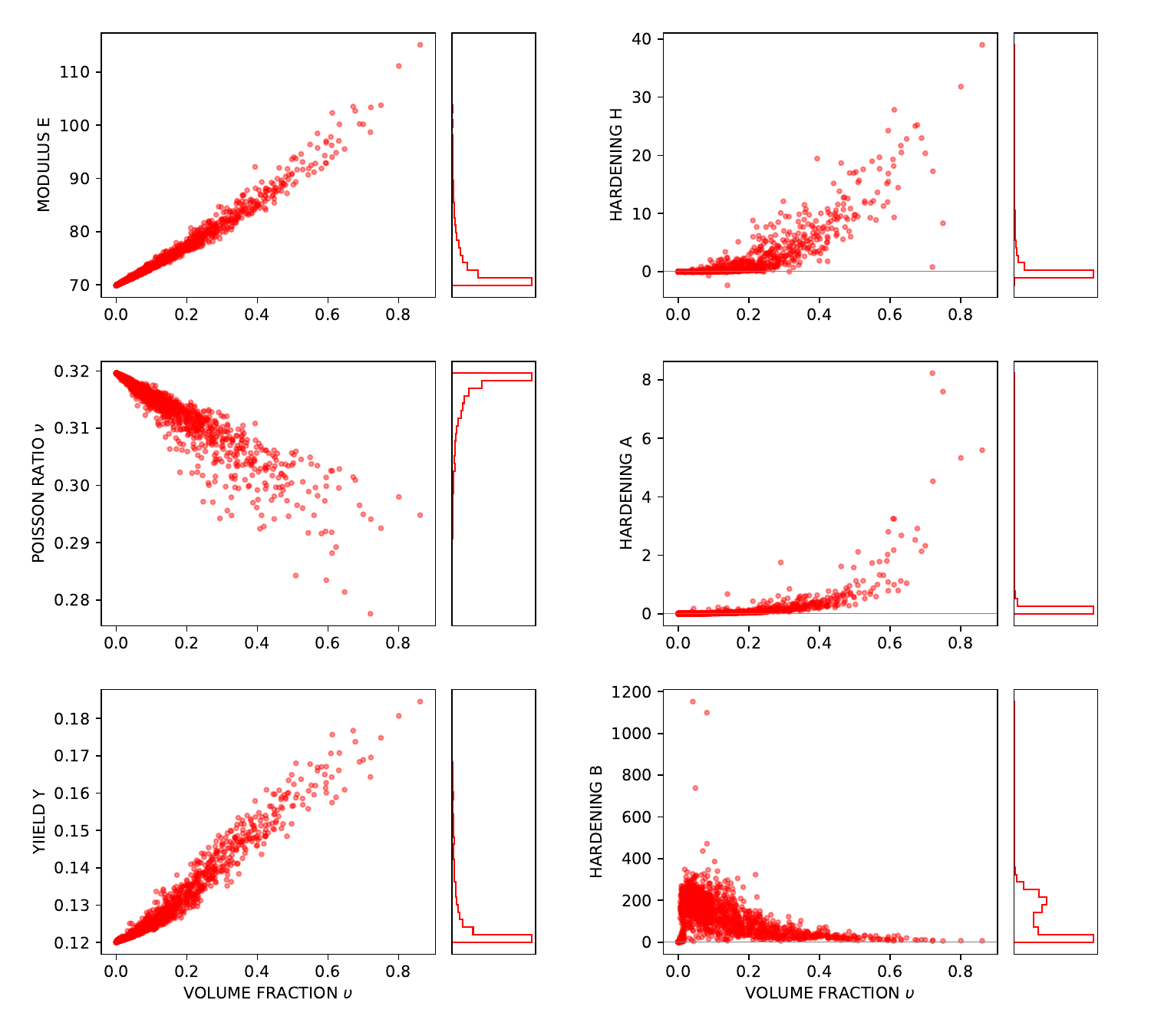}

\caption{Property correlations with volume fraction $\volfrac$.
Note stress-like quantities are in MPa.
}
\label{fig:property_correlations}
\end{figure}

To illustrate a case where there is significant variance, and spatial correlation, but not the extremes illustrated by the smaller SVEs we selected an SVE size of 24$^3$ elements since this was also a good trade-off between structure resolution and computational expense.
As mentioned, the SVEs are taken at random from multiple 10$^3$ larger realizations with 0.1 volume fraction of Si.
\fref{fig:realizations} shows representative SVEs for a range of volume fractions and their corresponding deformations and equivalent plastic strain fields.
As can be seen, there is a complex interplay of the structure and the imposed deformation.
Properties extracted from the 2000 SVEs used for training are summarized in \tref{tab:sve_properties}.
\fref{fig:property_correlations} shows property correlations with volume fraction.
Clearly, the response depends on structural aspects beyond just volume fraction, and a simple mixture model of the properties for Si and Al would be insufficient.
Also apparent are the effects of fitting to a model that has multiple capacities to match nearly linear behavior, particularly in the $B$ parameter.

\begin{figure}[htb!]
\centering
\includegraphics[width=0.60\textwidth]{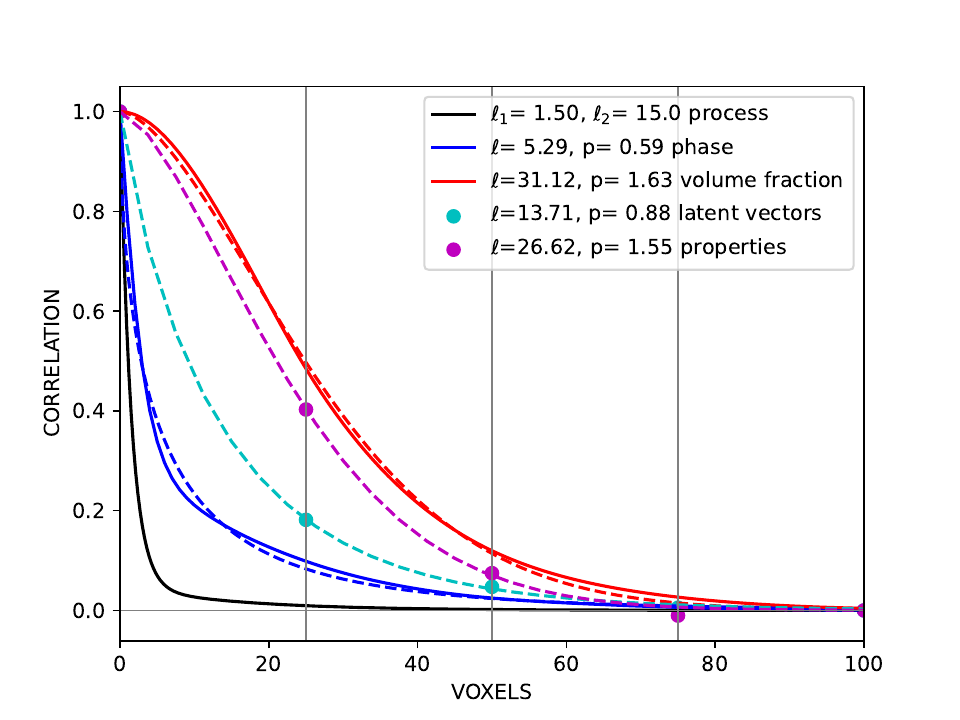}
\caption{Correlation of underlying process, resulting phase field, volume fraction of SVEs, as well as SVE latent vectors, and properties.
Correlation  lengths $\ell$ and decay exponents $p$ were extracted by fitting to $\exp(-(r/\ell)^p)$.
Note the minor negative correlations are likely due to numerical noise from undersampling at long correlation lengths.
}
\label{fig:correlation_lengths}
\end{figure}

\fref{fig:correlation_lengths} shows the correlation lengths of various structural-property aspects of the particular microstructural process we are using.
The underlying continuous structure-generating process has a significantly shorter correlation length than the binary voxel phase assignments.
The induced structural property volume fraction $\volumefraction$ and the mechanical properties of the SVEs have even longer correlation lengths.
The latent vectors generated by the NN encoder of the structure $\NN_\latentvector$, described in \sref{sec:architecture}, have a correlation length between that of the structure process and the induced property process.

\begin{table}
\centering
\begin{tabular}{|l|c|rl|}
\hline
{\bf Al} && & \\
\hline
Young's modulus & $E$  & 70.0 & GPa  \\
Poisson's ratio & $\nu$  & 0.32 &  \\
Yield strength  & $Y$ &  120 & MPa \\
Voce linear hardening coefficient & $H$ & 290.8 & MPa \\
Voce exponential hardening coefficient & $A$ & 1.0 & MPa \\
Voce exponential hardening prefactor & $B$ & 434.9 & \\
\hline
\hline
{\bf Si} && & \\
\hline
Young's modulus & $E$    & 140 & GPa   \\
Poisson's ratio & $\nu$  & 0.30 &  \\
\hline
\end{tabular}
\caption{Direct numerical simulations (DNS) properties
for AL and Si assumed elastic over range of interest.
}
\label{tab:dns_properties}
\end{table}

\begin{table}
\centering
\begin{tabular}{|l|c|rrrl|}
\hline
{\bf Property} & &min&max& std.error & \\
\hline
Volume fraction                   &  $\volfrac$ & 0.0  & 0.86  & 0.0000 &  \\
Young's moduli                    &  $E$        & 69.9 & 115.1 & 0.0299 & GPa \\
Poisson's ratio                   &  $\nu$      & 0.27 & 0.32  & 0.0000 &  \\
Yield strength                    &  $Y$        & 120  & 185   & 0.0005 & MPa \\
Linear hardening coefficient      & $H$         & 0.0  & 39.0  & 0.0937 & GPa \\
Exponential hardening coefficient & $A$         & 0.6  & 8226  & 0.0300 & MPa \\
Exponential hardening exponent    & $B$         & 0.84 & 1152  & 34.1342 & \\
\hline
\end{tabular}
\caption{Stochastic volume element (SVE) property ranges for nominal volume fraction $\volfrac=0.10$.
Overall model fit RMSE = 0.0023 MPa.
}
\label{tab:sve_properties}
\end{table}

\section{Structure-property encoding network} \label{sec:architecture}

\begin{figure}[htb!]
\centering
\includegraphics[width=0.65\textwidth]{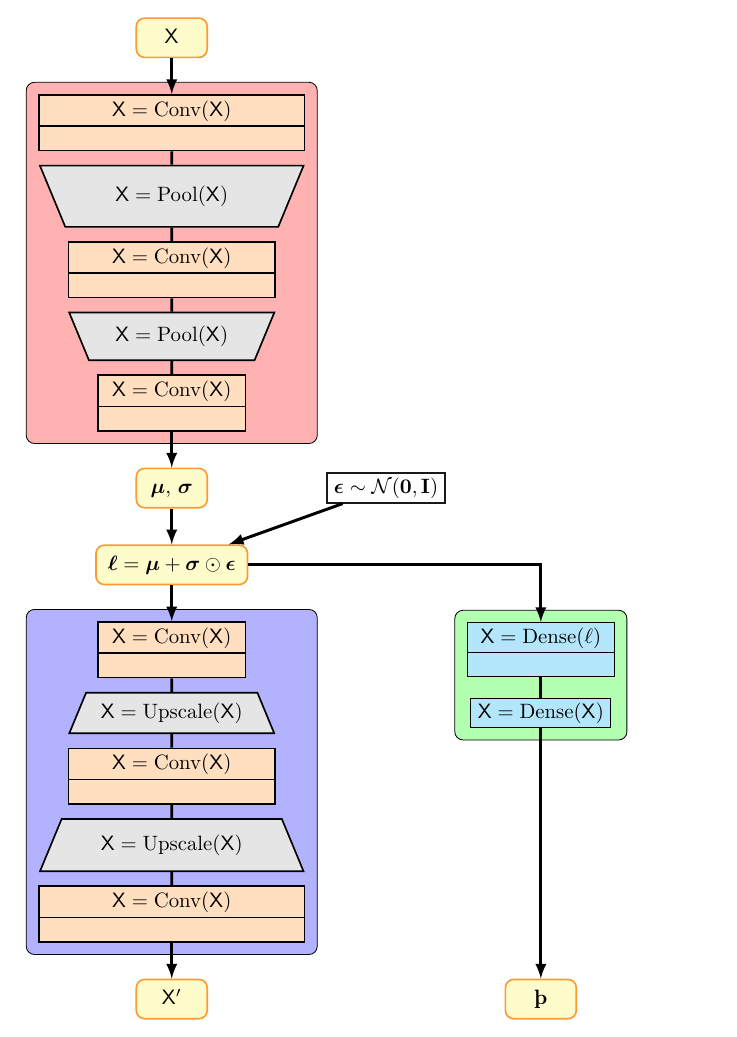}
\caption{Property variational autoencoder (pVAE) schematic.
The encoder (red) encodes the microstructure $\Xs=\phase(\xb)$ to a compact latent description $\latentvector$ which is then passed to the decoder (blue) to output a reconstruction $\Xs'$ and the property regressor (green) which outputs a property prediction $\propertyvector$.
The latent vector $\latentvector$ has a parametrized distribution $\Nc(\mub,\sigmab)$.
}
\label{fig:architecture_schematic}
\end{figure}

The neural network (NN) we use to encode the microstructures into a compact latent space is a variational autoencoder (VAE) \cite{kingma2019introduction} augmented with a regressor.
Since our ultimate goal is to encode the structure-to-property map we use an architecture akin to that of Wang \etal  \cite{wang2020deep}, which we will call a property variational autoencoder (pVAE).
This NN has three parts: (a) an encoder of the structure to latent representation $\phase \to \latentvector$, (b) decoder that takes latent representations back to structures $\latentvector \to \phase$, and (c) a regressor that takes latents to properties of the structure $\latentvector \to \propertyvector$.
The architecture is illustrated in the schematic in \fref{fig:architecture_schematic}.
Here $\phase(\xb)$ is the field describing the microstructure, in the present context the Al/Si phase, $\phase'(\xb)$ is its reconstruction, $\latentvector$ is the latent vector that encodes $\phase$, and $\propertyvector$ is the collection of properties associated with $\phase$.

The loss function
\begin{equation}
\Lc(\theta_e,\theta_d,\vartheta; \phase, \propertyvector)
= \Exp_{q_{\theta_e}(\latentvector|\phase)} \left[ \log p_{\theta_d}(\phase) \right]
- \beta \, \Exp_{q_{\theta_e}(\latentvector|\phase)} \left[ \log \frac{q_{\theta_e} (\latentvector | \phase)} { p_{\theta_d} (\latentvector | \phase ) } \right]
+ \lambda \, \| \propertyvector - \propertyvector_\vartheta(\latentvector) \|^2
\end{equation}
reflects the scalarization of three competing objectives:
(a) the data fit of the structure reconstruction $\phase \to \phase'$ from encoder to latent space to decoder,
(b) the Kullback-Leibler (KL) divergence between the latent distribution of the data $p_{\theta_d}$ and its representation by a surrogate distribution $q_\vartheta$, and (c) the regression error of the latent to property map $\propertyvector_\vartheta(\latentvector)$.
Here $q_{\theta_e}(\latentvector|\phase)$ is the distribution of latent vectors given microstructures produced by the encoder, and $p_{\theta_d} (\latentvector | \phase )$, conversely, is the distribution produced by the decoder.
The KL term promotes similiarity of the trained latent distribution and the distribution implicit in the data.
As in the original VAE we assume that the distribution $q_{\theta_e}(\latentvector|\phase) = \Nc(\mub,\sigmab)$ is well approximated by a multivariate, independent Gaussian distribution so that $\latentvector = \mub + \sigmab \odot \varepsilonb$ where $\odot$ is the component-wise Hadamard product and $\varepsilonb \sim \Nc(\mub,\mathbf{1})$.
The vectors $\latentvector$, $\mub$, $\sigmab$, and $\varepsilonb$ are all of length $N_\latentvector$, which is the size of the latent space and a hyperparameter.
The parameters $\mub$ and $\sigmab$ are trained at the same time as $\theta_e$, $\theta_d$, and $\vartheta$.
This assumption also allows for a particularly simple form of the KL divergence.
As in the $\beta$-VAE \cite{higgins2017beta} we also use hyperparameters $\beta$ and $\lambda$ to balance the relative scale and importance of the separate objectives.
In addition, these allow annealing, in the sense of a schedule of increasing and decreasing $\beta$ or $\lambda$, to aid training to the available data.
Reconstruction loss is mainly a regularizer of the regression task and is said to organize the latent space \cite{wang2020deep} by promoting it to be continuous with respect to the regression task.

The trainable parameters are summarized in \tref{tab:architecture_configuration}.
We exploited the relative computational cost of obtaining the data in training the pVAE network.
Since the encoder-decoder needed to be highly parameterized to encode the wide variety of structures, we first trained the NN to a large dataset of 100,000 samples of the Al-Si microstructure.
For this first stage the only property we expect the regressor to predict is the volume fraction since this data is obtainable with negligible cost.
Then we warm-started the network with a much smaller training set of 2000 samples where we added the 6 mechanical properties that were relatively expensive to compute.
Note that only the output layer of the regressor changed size in this stage.
After allowing the latent space to adjust to the new tasks, we increased $\lambda$ to make the training focus on the accuracy of the regressor.
(This could also be accomplished by freezing the parameters of the encoder-decoder.)

\begin{figure}[htb!]
\centering
\includegraphics[width=0.30\textwidth]{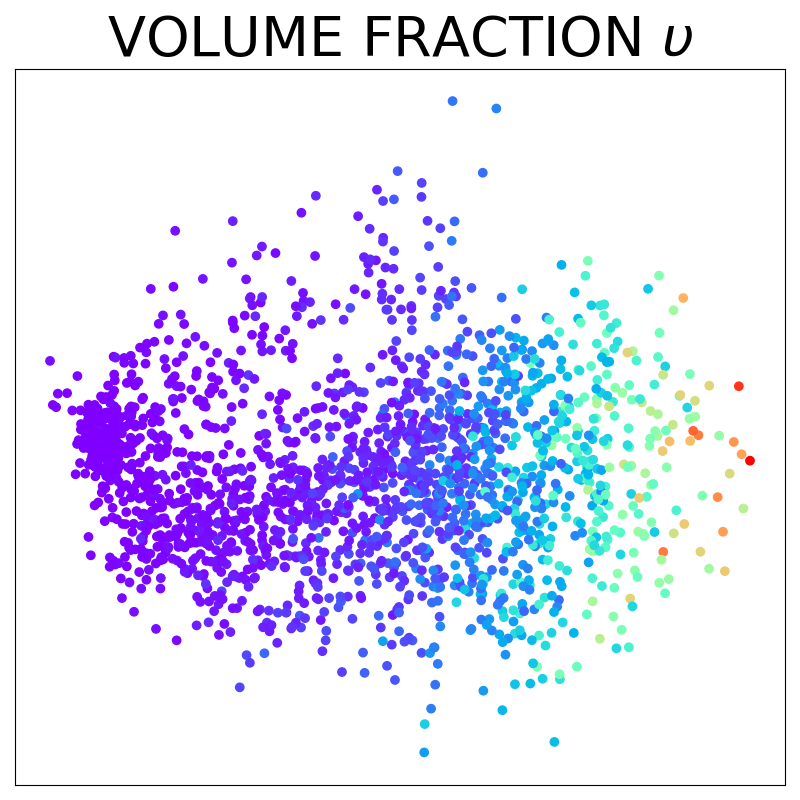}
\includegraphics[width=0.30\textwidth]{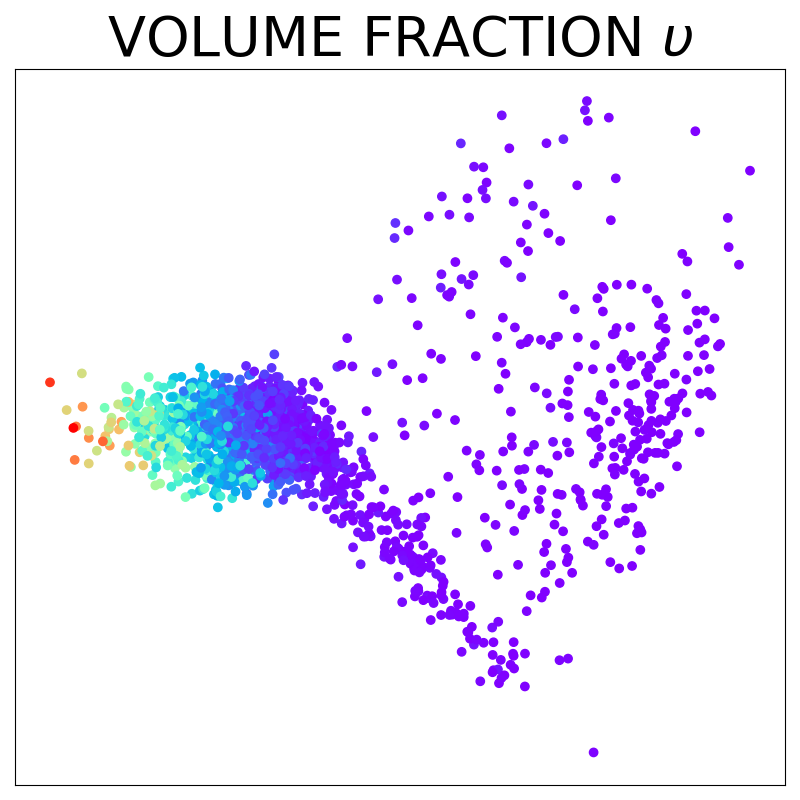}
\caption{Latent sample: first two PCA components colored by volume fraction: (left) pVAE trained to volume fraction only, (right) pVAE trained to all properties.
}
\label{fig:latent_volfrac}
\end{figure}

\begin{figure}[htb!]
\centering
\includegraphics[width=0.75\textwidth]{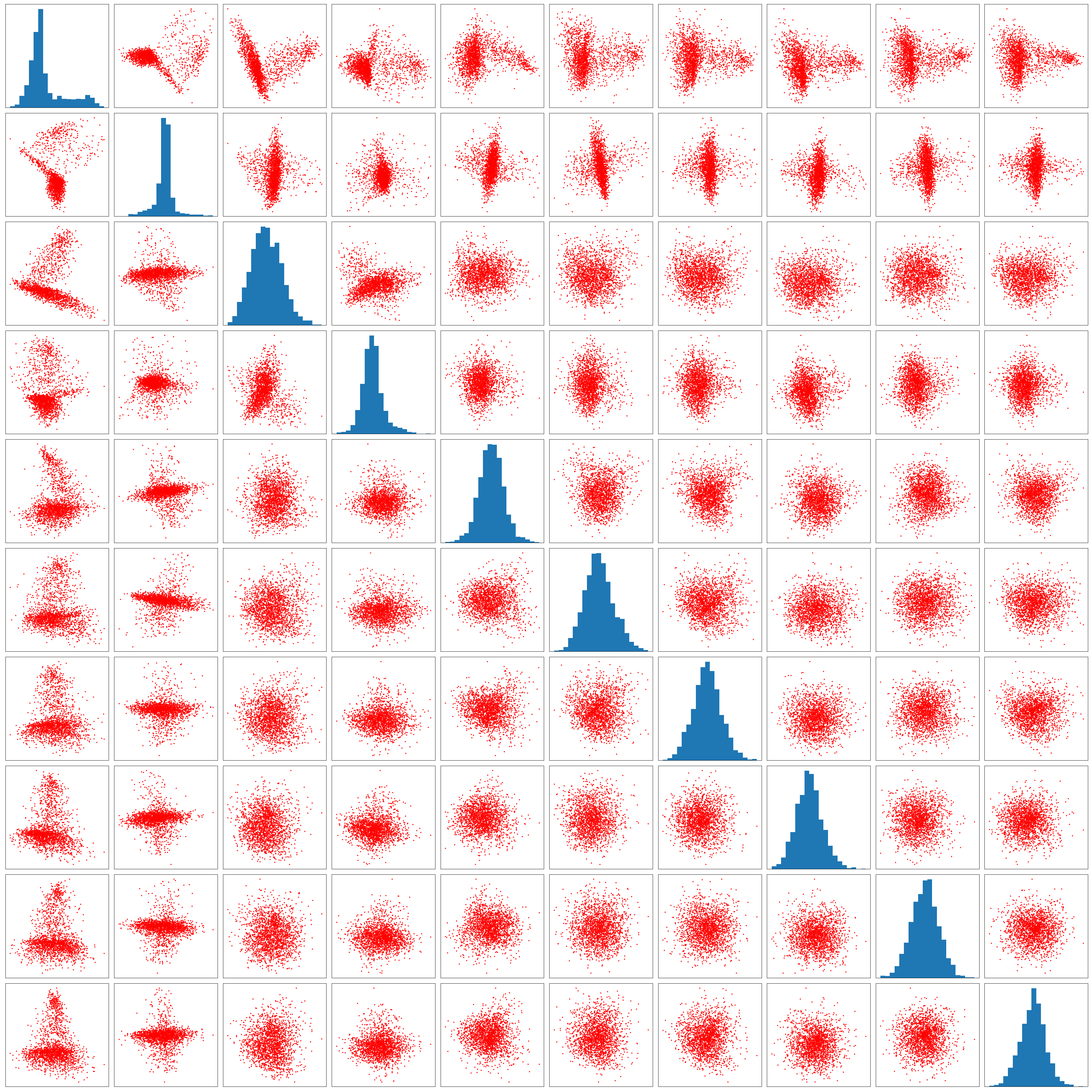}
\caption{Latent PCA components.
Note only 10 of the 20 components are shown for clarity.
}
\label{fig:latent_correlations}
\end{figure}

\fref{fig:latent_volfrac} shows how the latent space adapts to the new task of predicting the mechanical properties.
The latent samples are projected into the first two principal component analysis (PCA) components.
Clearly the latent space transitions from a generally Gaussian distribution to a distribution that is less normal.
\fref{fig:latent_correlations} gives projected views in the first 10 PCA components.
The distribution in the first four components appear to be non-Gaussian, while the remaining, less significant components appear to be reasonably Gaussian.

\begin{figure}[htb!]
\centering
\includegraphics[width=0.45\textwidth]{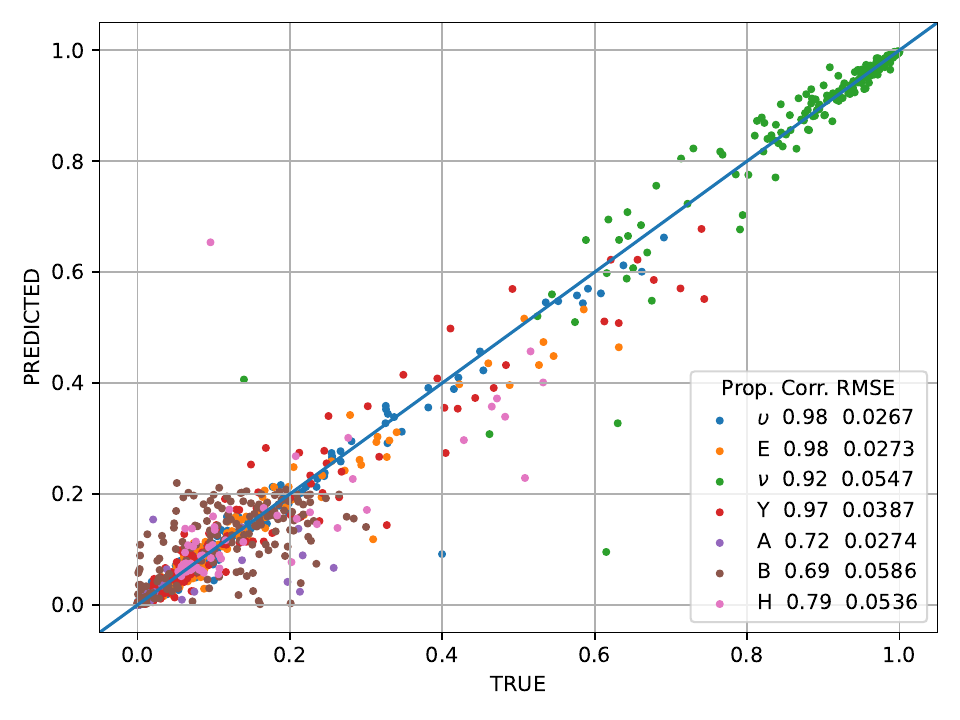}
\includegraphics[width=0.50\textwidth]{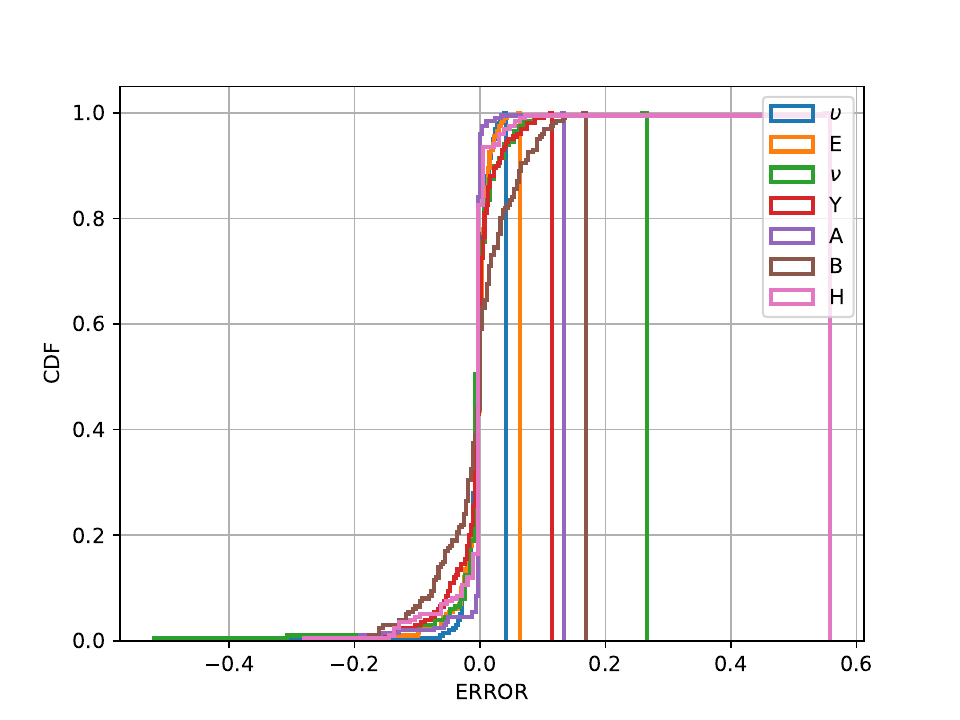}
\caption{Property prediction correlations with held out data (left) and CDF of relative/normalized errors (right).
Note the CDFs terminate at the highest sample in the data.
}
\label{fig:property_error_cdfs}
\end{figure}

\fref{fig:property_error_cdfs} shows that the property regression errors for the final pVAE are generally between 2 and 5 percent.
The properties with more complex distributions and less simple correlations with the SVE volume fraction, namely the hardening parameters $H$ and $B$, have the highest errors.
We attribute the high relative error in the regression of the Poisson's ratio to the narrow range of the true values  (Al and Si have similar $\nu$) and hence the low sensitivity of the Poisson's ratio to the structure.

\begin{table}
\centering
\begin{tabular}{|ccc|}
\hline
Layer & output & parameters \\
\hline
\hline
{\bf Encoder} && \\
\hline
convolution& 24$^3$ $\times$ 16 &  448  \\
max pooling& 12$^3$ $\times$ 16 &  0 \\
convolution& 12$^3$ $\times$ 32 &  13856  \\
max pooling&  6$^3$ $\times$ 32 &  0 \\
convolution&  6$^3$ $\times$ 64 &  55360  \\
\hline
\hline
{\bf Latent} && \\
dense & 2 $\times$ $N_\latentvector$&  840 \\
\hline
\hline
{\bf Decoder} && \\
convolution& 12$^3$ $\times$ 32 &  55328  \\
convolution& 24$^3$ $\times$ 16 &  13840  \\
convolution& 24$^3$ $\times$ 1  &  433  \\
\hline
\hline
{\bf Regressor} && \\
dense&  $N_\latentvector$ &  420 \\
dense&  $N_\latentvector$ &  420 \\
dense&  $N_\latentvector$ &  420 \\
dense&  $N_\latentvector$ &  420 \\
dense&  $N_\propertyvector$ &  147 \\
\hline
\end{tabular}
\caption{Property variational autoencoder (pVAE) configuration.
GELUs were used for all activations including the property output layer since all properties are non-negative.
Dependence on batch size suppressed for clarity.
}
\label{tab:architecture_configuration}
\end{table}

\begin{figure}[htb!]
\centering
\includegraphics[width=0.75\textwidth]{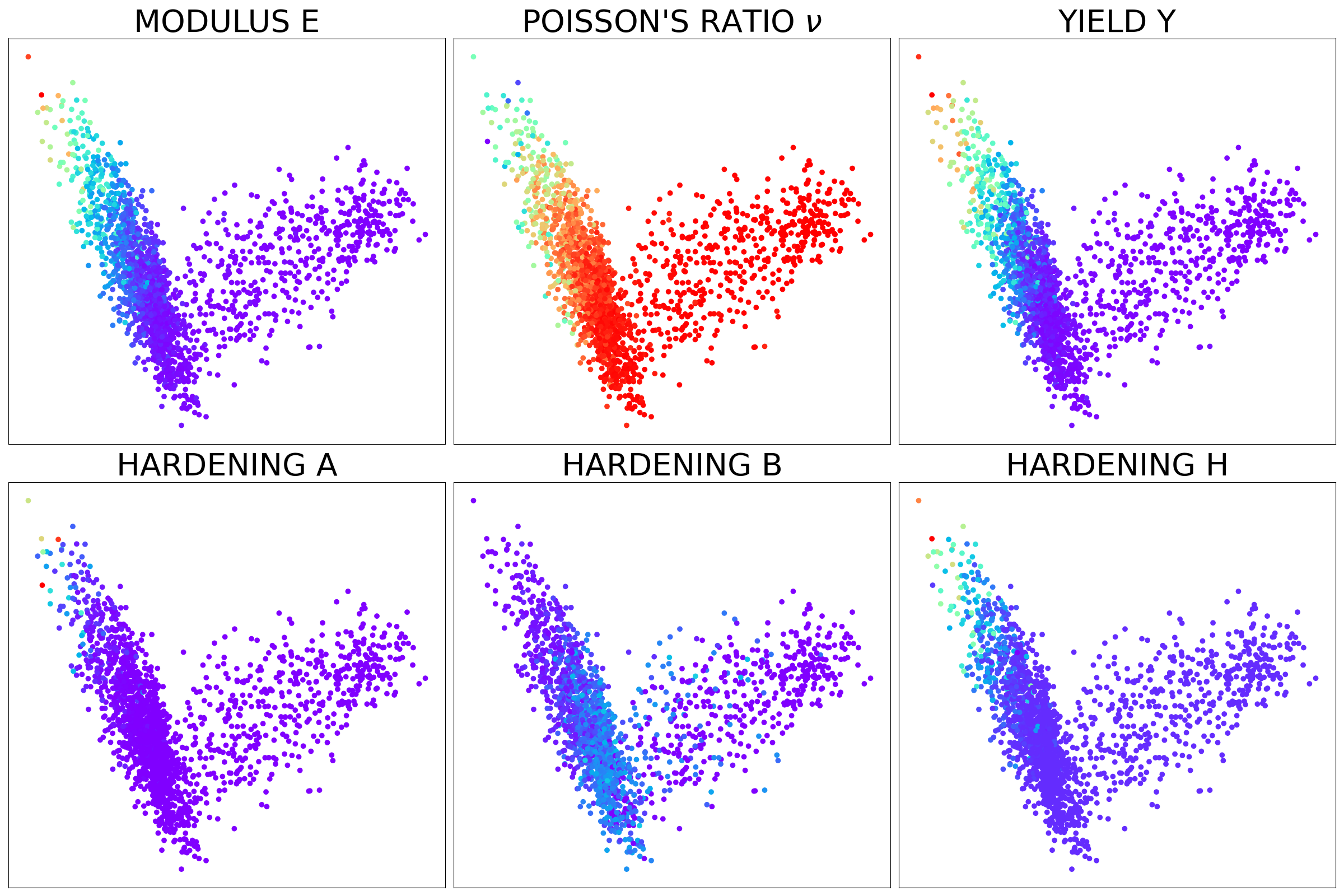}
\caption{Latent sample: first and third PCA components colored by corresponding properties indicating semantic directions.
}
\label{fig:latent-semantic}
\end{figure}

Latent encoding of properties serves the purpose of giving a compact representation of the correlation of properties at any point, and hence the structure-property correlations through the latent features.
\fref{fig:latent-semantic} shows the variation of each property with the first and third PCA components of the latent space, which are the most aligned with variations in properties.
Clearly, there is some correlation between all properties.
We conjecture that the second PCA component does not correlate significantly with properties since it is used to facilitate the structure reconstruction of the pVAE.
Also, note that more diffuse lobes in the scatter plots are low density and likely due to fitting the hardening with two competing functions, see \eref{eq:voce}.

\begin{figure}[htb!]
\centering
\includegraphics[width=0.65\textwidth]{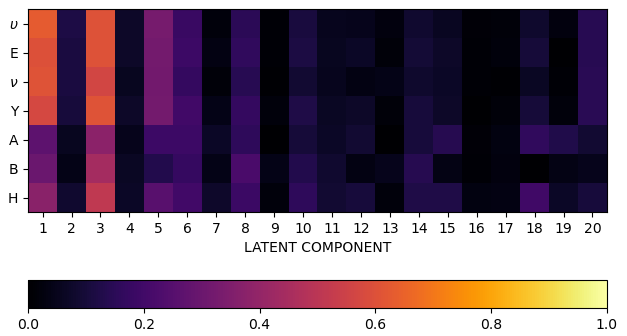}
\caption{Latent-property correlations in terms of PCA components.
}
\label{fig:latent-property_correlations}
\end{figure}

We associate the components of the \emph{semantic direction}  of a property $\property(\latentvector)$ with the correlation coefficient of the (orthogonal) PCA directions $\latentvector_I$
\begin{equation}
\semanticvector_I =
\frac{\langle \property, \latentvector_I \rangle}
{\sqrt{\langle \property, \property \rangle \langle \latentvector_I, \latentvector_I \rangle}}
\end{equation}
where $\langle \property, \property \rangle$ is the variance of $\property$.
The semantic direction, in the sense of the direction of highest variance (and gradient),  may not be constant in latent space with a complex locus of representative samples.
\fref{fig:latent-property_correlations} shows that generally all the properties are strongly correlated with the first and third PCA component of the latent space although the correlation is lower for the hardening parameters that have a more complex dependence.
There are also significant correlations of all properties with less significant PCA directions.
Clearly, there is no perfect, simple alignment of the latent space variance with the property variations.

\section{Demonstrations} \label{sec:demonstrations}

In this section, we provide: (a) a comparison of the proposed method with a DNS, (b) a demonstration of its use in propagating microstructural uncertainty to mechanical quantities of interest, and (c) an illustration of how it can be used to model materials with microstructural functional gradation.
Each demonstration utilizes a component geometry with three competing holes, which is meant to be representative of a topological optimization scenario with an uncertain material.
Each is loaded in tension for ease of interpretation by applying fixed $x_1$ boundary conditions on one end, $x_2$ and $x_3$ boundary conditions on edges of this end and prescribed boundary conditions on the opposing end.
For the DNS comparison, we used a 84$\times$48$\times$4 $\mu$m component, and for the generative demonstrations we used 240$\times$40$\times$4 $\mu$m  configurations.
Each multiscale simulation was performed with a standard finite element simulator \cite{beckwith2022sierra} that was modified to take the property regressor component of the trained pVAE and a field of latent vectors.
Since the imprinting of the mesh with a latent vector field is done as a preprocessing step and the complexity of the constitutive model is comparable to that of a traditional model, the speed up is proportional to the reduction in the number of elements, \ie for a DNS boundary value problem encoded into $n^3$-element SVEs the speed up is $n^3$.

\subsection{DNS comparison} \label{sec:dns}
First, we show how well the proposed method approximates a corresponding DNS.
Using the GRF structure generation algorithm we created a large 504$\times$288$\times$24 voxel structure.
A structured rectangular grid was employed to facilitate partitioning the overall structure into disjoint SVEs.
Using the trained pVAE we encoded each of the SVEs into latent vectors that will be subsequently used in a multiscale simulation and mapped to properties.
This DNS was compared to a corresponding multiscale simulation with a selected SVE size.

\begin{figure}[htb!]
\centering
\includegraphics[width=0.55\textwidth]{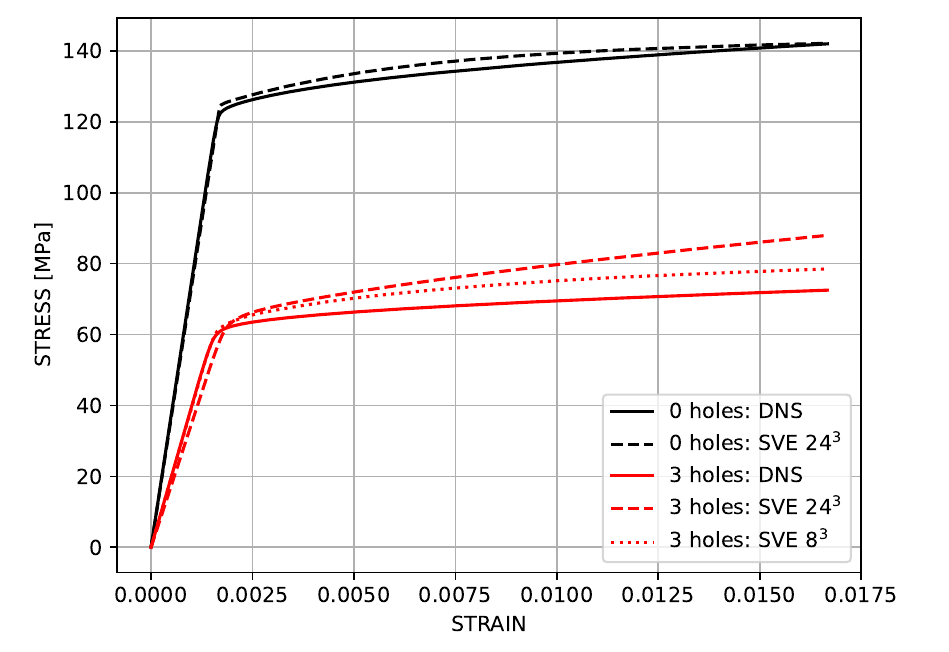}
\caption{DNS comparison of sample norminal stress-strain response.
Nominal strain was obtained from the end displacement of the sample and nominal strain was calculated from the reaction force on the end and its initial area.
}
\label{fig:dns_comp_response}
\end{figure}

\begin{figure}[htb!]
\centering
\includegraphics[width=0.55\textwidth]{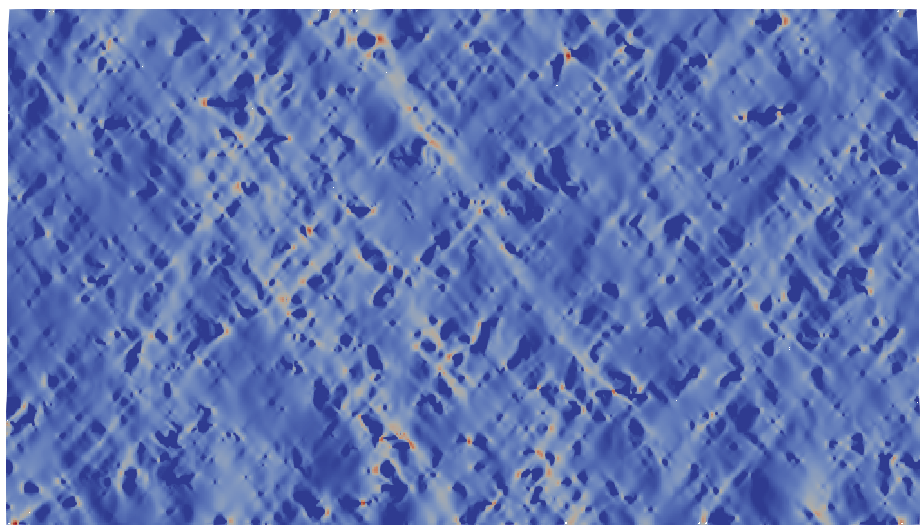}
\includegraphics[width=0.55\textwidth]{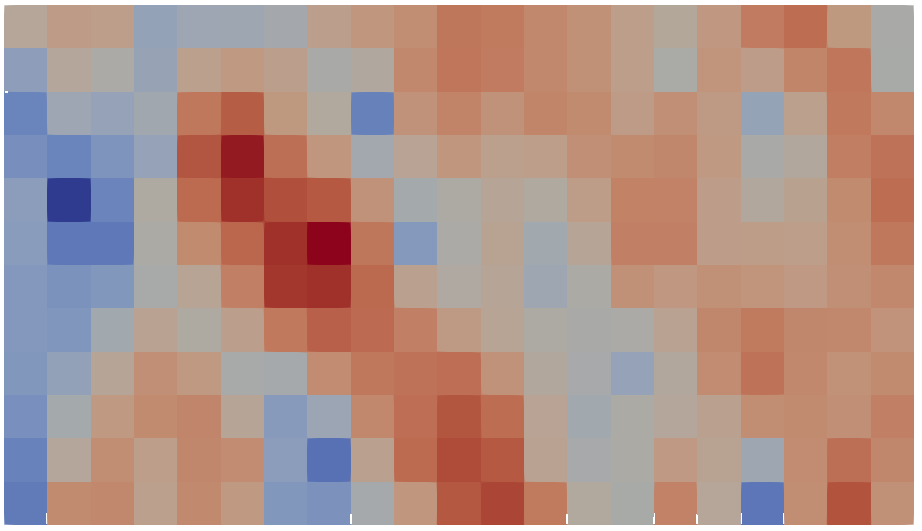}
\caption{DNS comparison of localization for the part without holes.
Equivalent plastic strain at nominal strain 0.007,
blue: 0.0, red
DNS  0.028
SVE 24$^3$ 0.009,
}
\label{fig:dns_comp_nohole}
\end{figure}

\begin{figure}[htb!]
\centering
\includegraphics[width=0.45\textwidth]{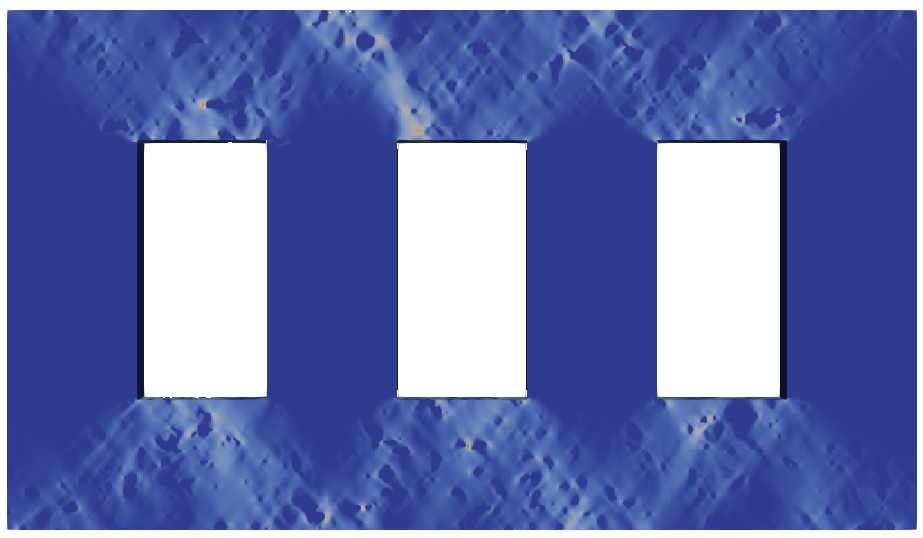}

\includegraphics[width=0.45\textwidth]{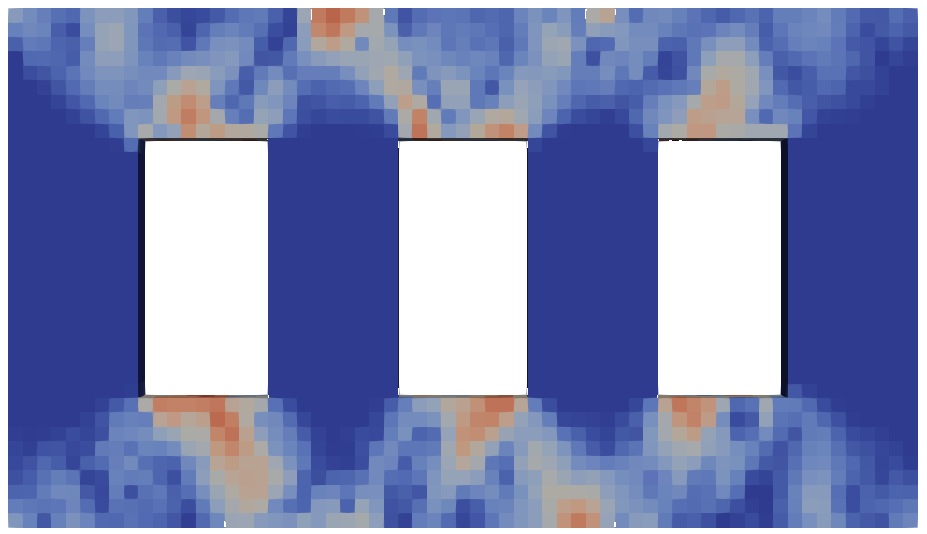}

\includegraphics[width=0.45\textwidth]{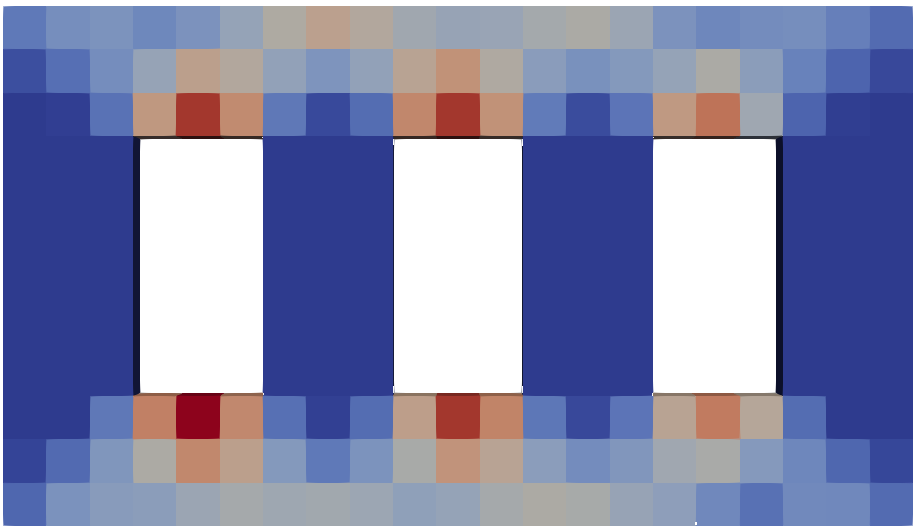}
\caption{DNS comparison of localization for part with 3 holes.
Equivalent plastic strain at nominal strain 0.007,
blue: 0.0, red
DNS  0.069,
SVE 24$^3$ 0.020,
SVE  8$^3$ 0.028.
}
\label{fig:dns_comp_hole}
\end{figure}

\fref{fig:dns_comp_response} shows that if there are no holes, as in \fref{fig:dns_comp_nohole}, there are weak gradients and the multiscale method approximates the response of the DNS.
\fref{fig:dns_comp_nohole} also shows that the coarse field resembles that of the DNS in this case.
Holes, as in the configuration shown in \fref{fig:dns_comp_hole}, create gradients that interact with the microstructure.
There is apparently a bias in the hardening response that leads the multiscale methods to overestimate the nominal stress in this scenario.
In this case, the multiscale response with a coarse 24$^3$ SVEs is moderately far from the DNS response; however, employing 8$^3$ SVEs provides better resolution of the gradients and brings the response closer to that of the DNS.


\subsection{Uncertainty quantification} \label{sec:uq}

To propagate uncertainty in performance quantities of interest (QoI) due to uncertainty in the material microstructure, we construct an ensemble of realizations of the latent fields on a mesh of the component and then evaluate the resulting ensemble of responses.
To construct realizations of the latent field $\latentvector(\xb)$ consistent with the chosen microstructure we need a method that embeds the spatial correlations of the latent space and the point-wise latent component correlations that embed the correlations between properties.
A KLE is adaptable to this purpose.
A KLE uses a kernel to construct a (deterministic) correlation matrix for the mesh by encoding spatial interactions  via
\begin{equation}
\Sigmab_{IJ} = \correlationkernel(\xb_I,\xb_J)
= \correlationkernel(\| \xb_I-\xb_J \|)
\end{equation}
by taking $\xb_I$ to be cell/element centroids of the mesh employed for the realization.
The corresponding eigenvectors $\vb_a$ and eigenvalues $\lambda_a$ of $\Sigmab$ are then used in the expansion:
\begin{equation} \label{eq:kle_realization}
\varphi(\xb) = \sum_a \eta_a \sqrt{\lambda_a} \vb_a
\end{equation}
which is truncated so that the largest included eigenvalues represent a given fraction of the total variance (we used 0.99).
Here, $\eta$ is a uncorrelated noise process, e.g. $\eta \sim \Nc(0,1)$, with realizations $\eta_a$.
This construction is sufficient to represent the spatial correlation of a scalar random field.
To account for the point-wise correlations of $\latentvector$ we estimated the correlation $C_{ab} = \langle \latent_a, \latent_b \rangle$ from encoding the large training sample and then construct:
\begin{equation} \label{eq:kle_vector_realization}
\latentvector(\xb) = \sum_a \eta_a \sqrt{\lambda_a} \Ls \vb_a
\end{equation}
using the Cholesky decomposition $\Ls$ of $\Cs  = \Ls \Ls^T$ \cite{daw2022overview}.
Modifications of this representation exist for non-Gaussian fields \cite{phoon2005simulation}, but we assumed this representation was sufficient for the current context.
Also, the samples used in training are equivalent to an empirical distribution that could have been used for this purpose, but the route we chose is extensible to as many independent realizations as needed and provides some filtering of outliers.

\begin{figure}[htb!]
\centering
\includegraphics[width=0.65\textwidth]{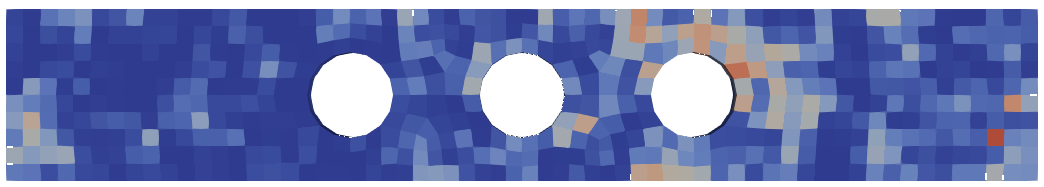}
\includegraphics[width=0.65\textwidth]{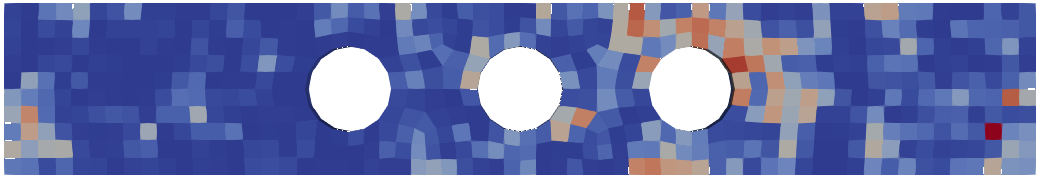}

\caption{Uncertainty quantification:
One realization of the bar with 3 holes.
Variation of (upper) modulus $E$ and (lower) yield strength $Y$.
The blue-red range of $E \in [69.9, 115.1]$ and $Y \in [120,185]$ are given in \tref{tab:sve_properties}.
}
\label{fig:uq_realization}
\end{figure}

\begin{figure}[htb!]
\centering
\includegraphics[width=0.45\textwidth]{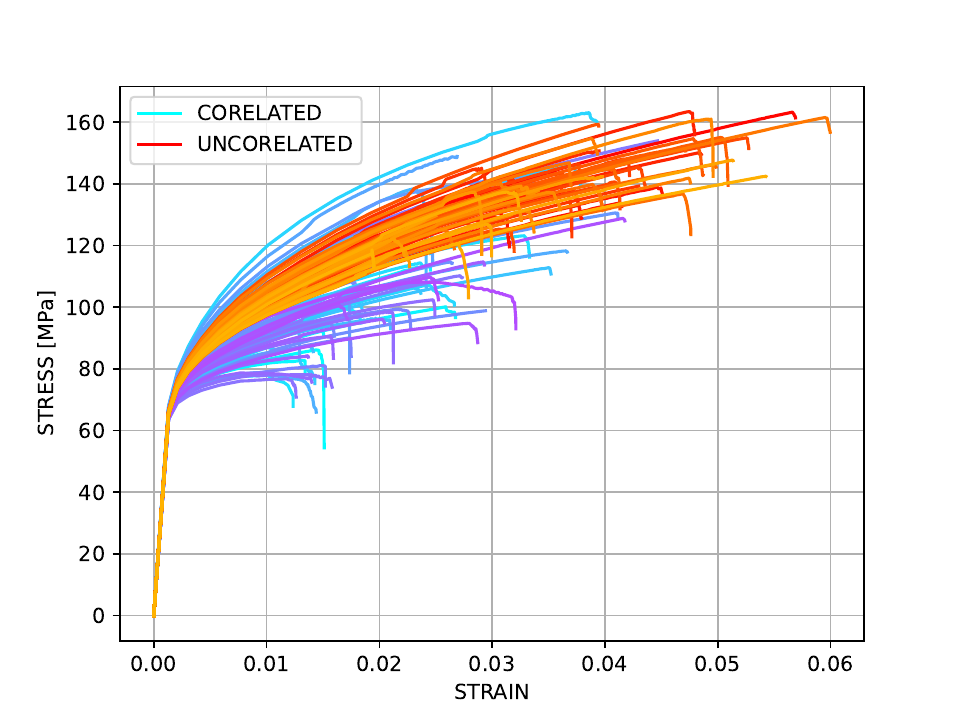}
\includegraphics[width=0.45\textwidth]{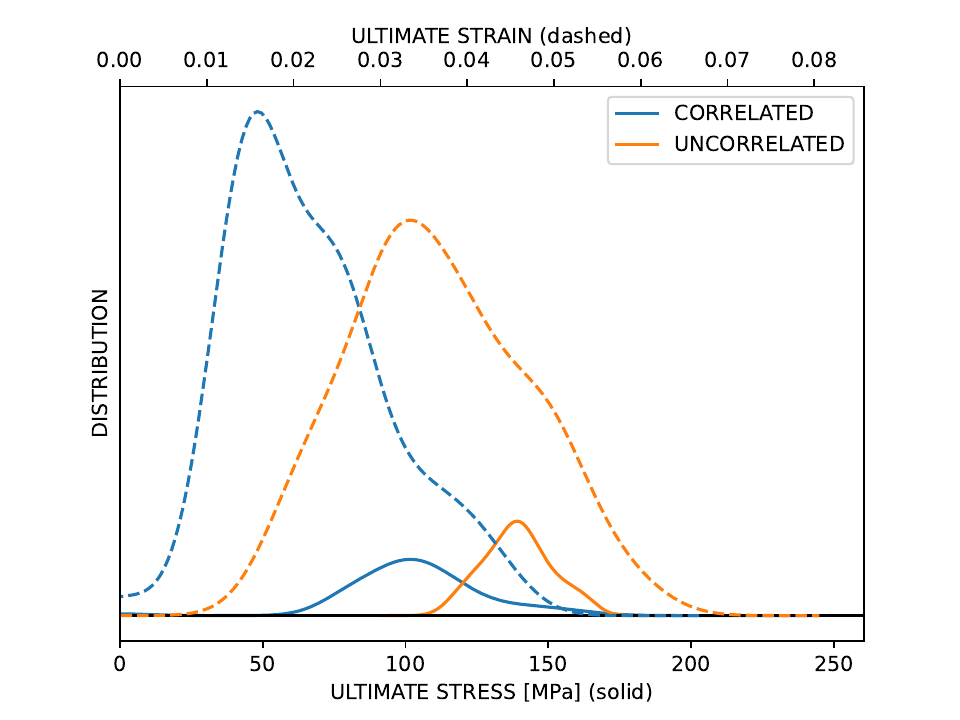}
\caption{Uncertainty quantification:
stress-strain response (left) uncorrelated and (right) spatially correlated realizations.
}
\label{fig:uq_response}
\end{figure}

\begin{figure}[htb!]
\centering
\includegraphics[width=0.65\textwidth]{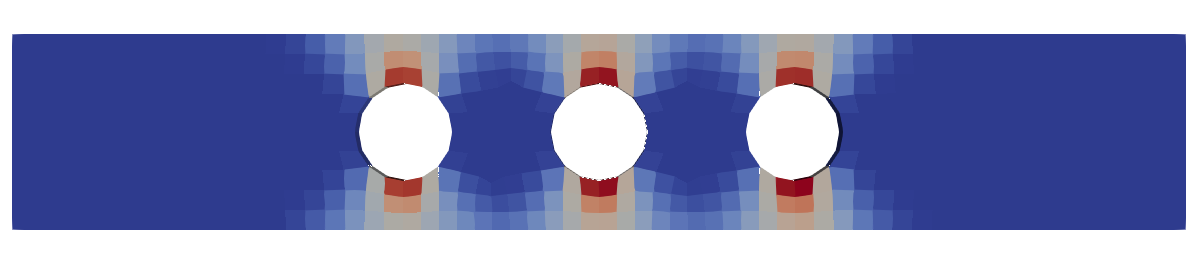}
\includegraphics[width=0.65\textwidth]{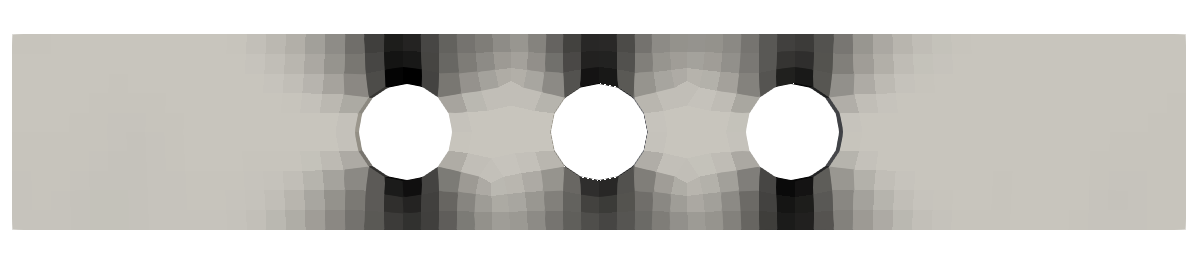}

\includegraphics[width=0.65\textwidth]{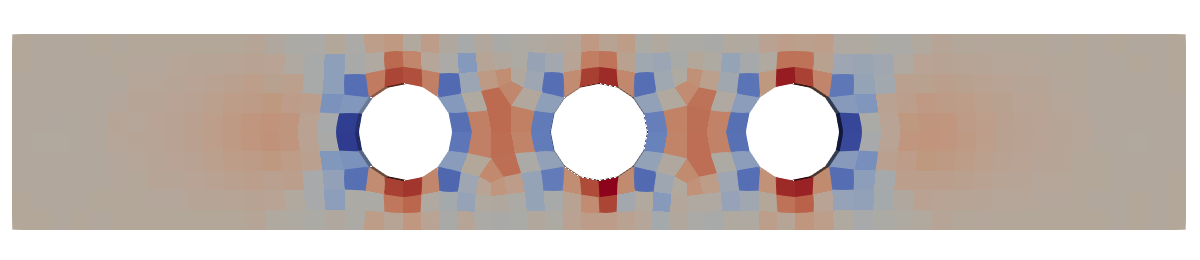}
\includegraphics[width=0.65\textwidth]{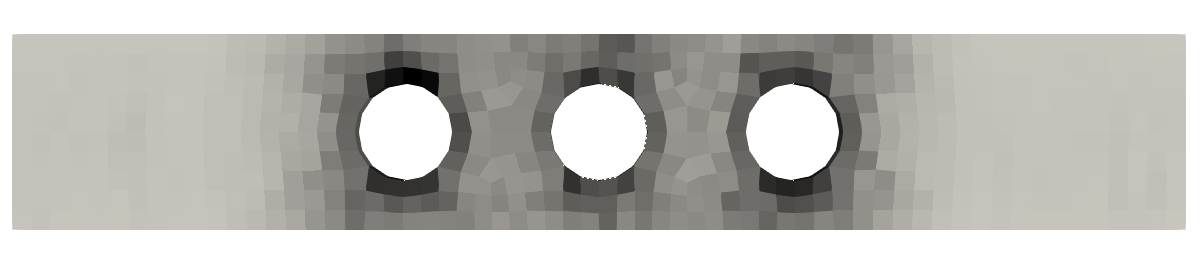}
\caption{Uncertainty quantification:
equivalent plastic strain (top) at nominal strain 0.007 (blue: 0.00, red:0.08, white:0.000, black:0.056)  and stress (bottom, blue: -85 MPa red:60 MPa, white:0 MPa, black:66 MPa) fields.
Mean field is in color and the standard deviation field is in gray.
}
\label{fig:uq_fields}
\end{figure}

To illustrate the effect of the spatial correlation across the realizations we simulated a scenario where the structure correlation in the latent vectors is comparable to deformation gradients and compare it to a corresponding one where the spatial correlation is too short to be effectively uncorrelated over the mesh.
\fref{fig:uq_realization} shows a realization of the correlated structure where the latents have been mapped to their corresponding properties.
\fref{fig:uq_response} compares the two response ensembles of 64 realizations each, interestingly the spatially correlated ensemble clearly has the higher variance apparently due to the fact that the spatial correlation reinforces high or low properties at the necks near the holes.
In contrast, the uncorrelated ensemble tends to have elements with moderate properties near elements with outlying properties.
\fref{fig:uq_fields} corroborates this conjecture.
The necks and near the interior holes have the highest equivalent plastic strain and stress, respectively, and also the highest variance in those critical fields.

\subsection{Functional gradation}

\begin{figure}[htb!]
\centering
\includegraphics[width=0.65\textwidth]{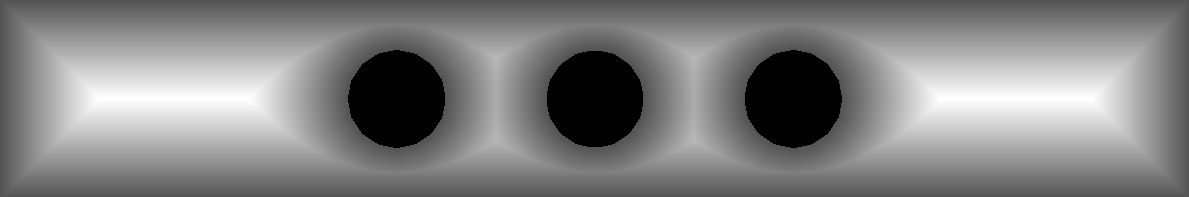}
\includegraphics[width=0.65\textwidth]{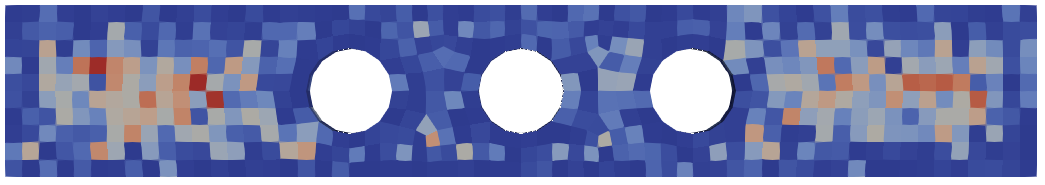}
\includegraphics[width=0.65\textwidth]{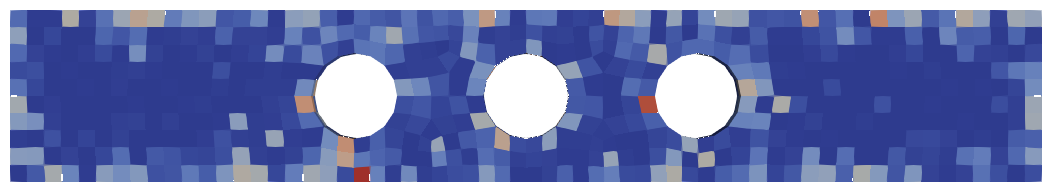}
\caption{Functional gradation:
(top) distance field.
(middle) yield stress $Y$  inward gradation,
(bottom) yield stress $Y$ outward gradation.
}
\label{fig:func_grad}
\end{figure}

Functional gradation is achieved by modulating the manufacturing process to create gradations of properties over a component.
To mimic this process, we use the semantic direction for volume fraction $\upsilon$ to create a non-stationary distribution
\begin{equation}
\pi(\latentvector; \xi) = \Nc(\mub+ \xi \, \sb, \Sigmab)
\end{equation}
which is a function of $\xi$ which we associate with the normalized distance to the surface, as shown in \fref{fig:func_grad}.
The semantic direction of the volume fraction was chosen since this is directly accessible to a manufacturing process, however the mechanical properties also vary with it.
We chose $\mu$ to be the center of the training latent space and $\Sigmab$ to be 0.5 of the covariance of the training latent space to represent a more tightly controlled process.

\begin{figure}[htb!]
\centering
\includegraphics[width=0.45\textwidth]{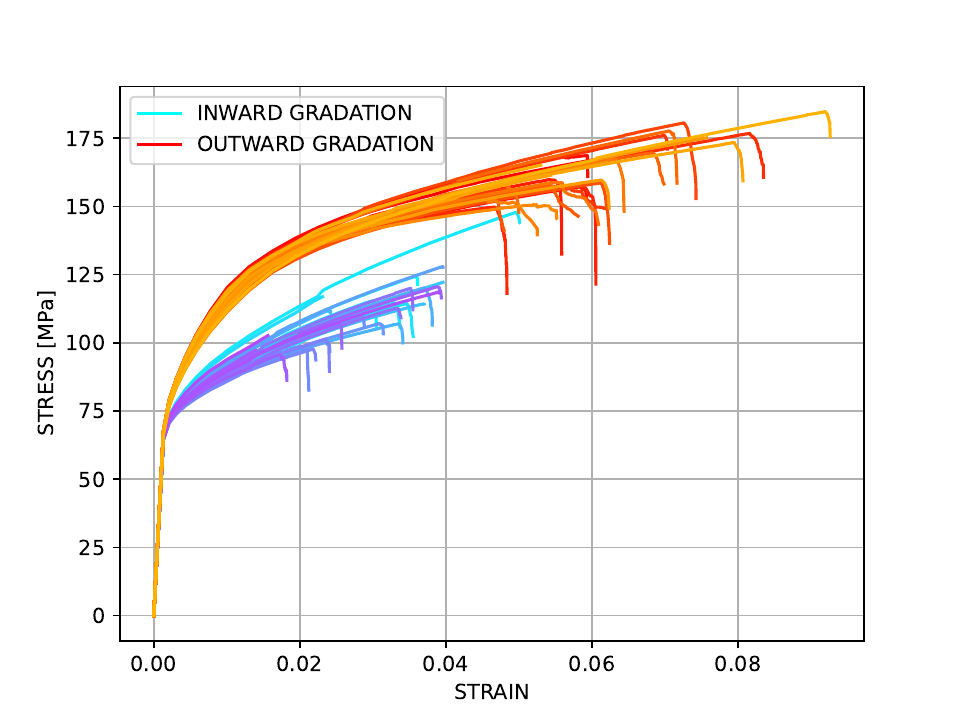}
\includegraphics[width=0.45\textwidth]{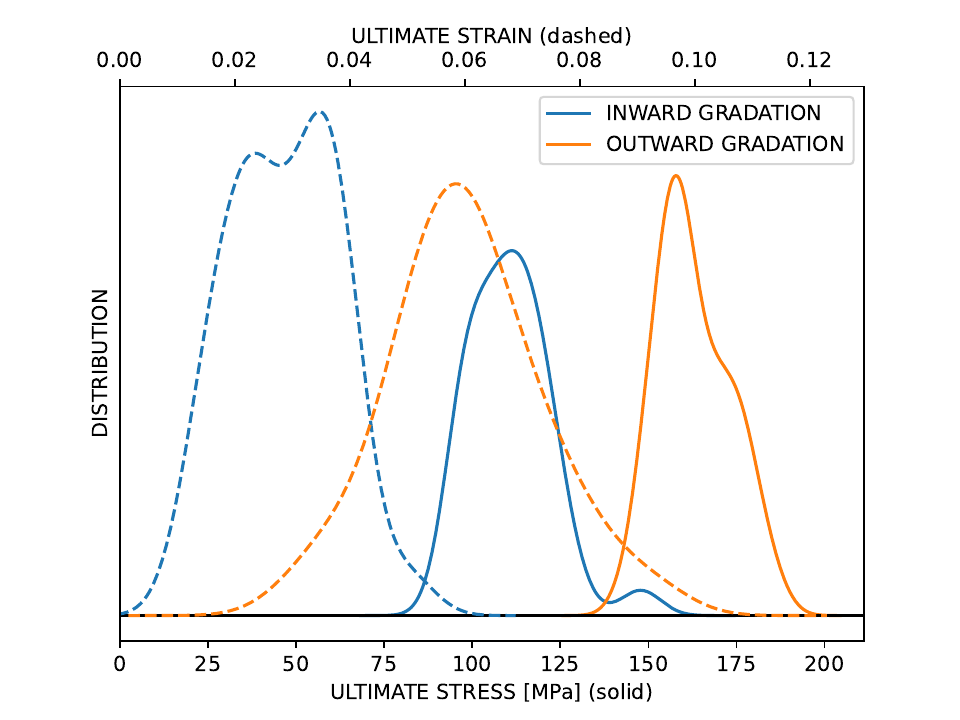}
\caption{Functional gradation:
stress-strain response inward (top) and outward (bottom) gradation.
}
\label{fig:fg_response}
\end{figure}

\begin{figure}[htb!]
\centering
\includegraphics[width=0.65\textwidth]{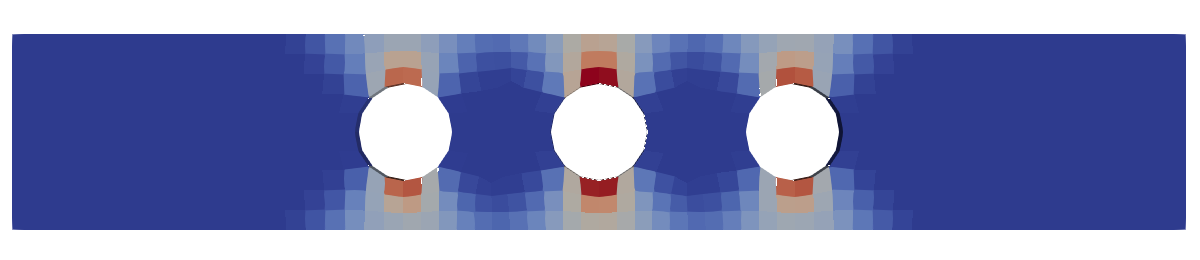}
\includegraphics[width=0.65\textwidth]{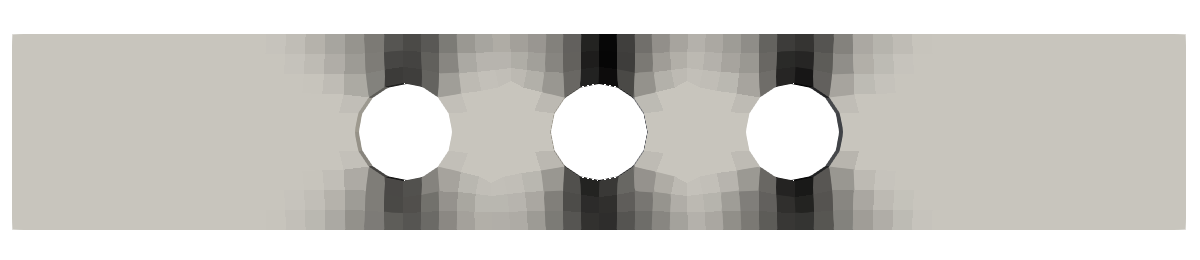}

\includegraphics[width=0.65\textwidth]{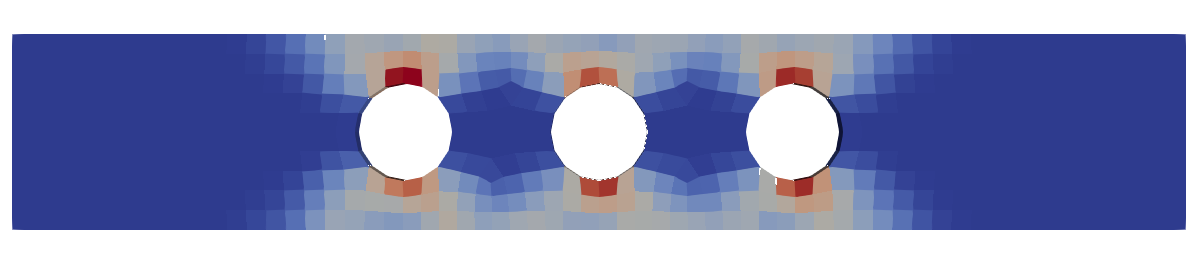}
\includegraphics[width=0.65\textwidth]{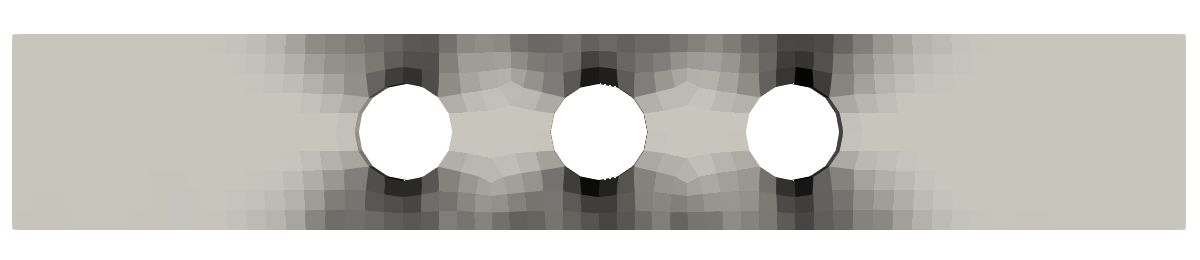}
\caption{Functional gradation: equivalent plastic strain field at nominal strain 0.007.
Inward (top) blue:0, red:0.075, white: 0.0, black:0.029.
Outward (bottom) blue:0, red:0.050, white: 0.0, black:0.022.
Mean (upper, color), standard deviation (lower, gray).
}
\label{fig:fg_eqps}
\end{figure}

\begin{figure}[htb!]
\centering
\includegraphics[width=0.65\textwidth]{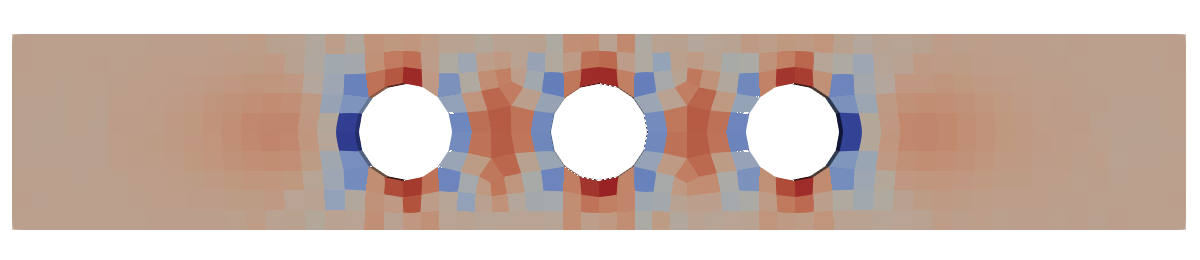}
\includegraphics[width=0.65\textwidth]{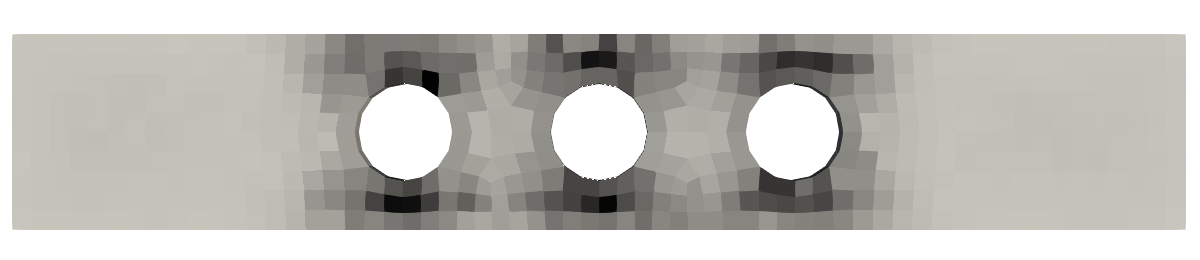}

\includegraphics[width=0.65\textwidth]{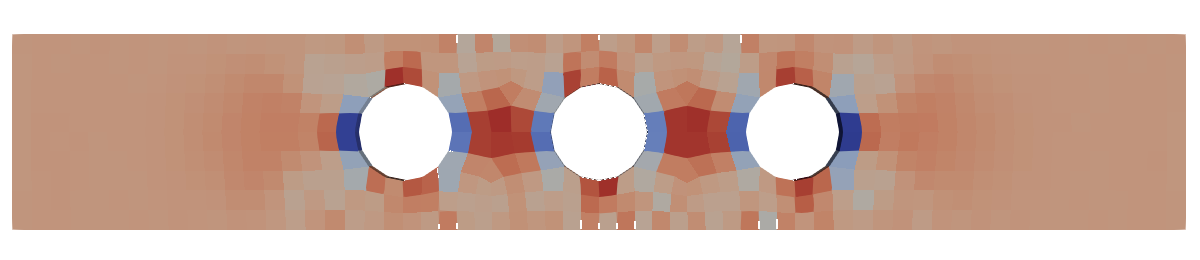}
\includegraphics[width=0.65\textwidth]{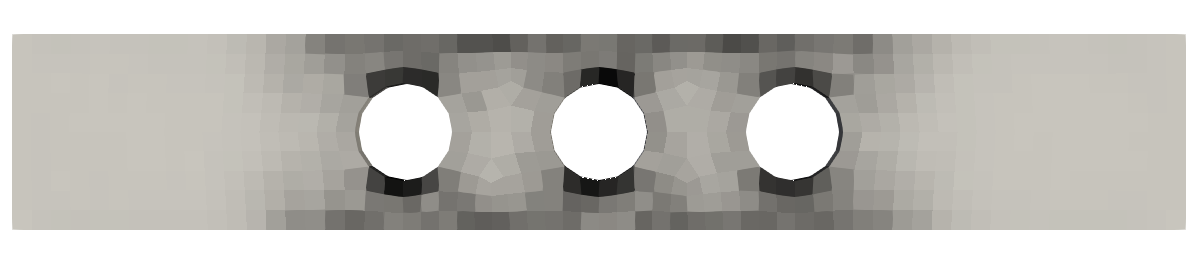}
\caption{Functional gradation: 11-stress field at nominal strain 0.007.
Inward (top) blue:-87 MPa, red: 63 MPa, white: 0.0, black:67 MPa.
Outward (bottom) blue:-11, red: 70 MPa, white: 0.0, black:11 MPa.
Mean (upper, color), standard deviation (lower, gray).
}
\label{fig:fg_stress}
\end{figure}

We compare two scenarios: (a) where the Si volume fraction increases toward the surface (outward gradient), and (b) where the Si volume fraction increases toward the interior of the part (inward gradient).
\fref{fig:func_grad} illustrates the functional gradation in properties for the inward scenario, where the trend is clear as is the stochastic variation.
\fref{fig:fg_response} shows that these two scenarios produce distinct ensembles of (32) responses and corresponding distributions performance QoIs.
Clearly, having higher concentration of Si near the surface leads to tougher parts.
\fref{fig:fg_eqps} and \fref{fig:fg_stress} illustrate that the local changes in response are fairly subtle.
The outward variation appears to spread the concentration of equivalent plastic strain and stress near the holes.

\section{Conclusion} \label{sec:conclusion}

Using a pVAE structure-property representation, KLE realization generation, and a standard finite element simulator we developed a multiscale method that accounts for the finite-size effects and spatial correlations that complicate simulation scenarios where there is no clear separation of the scales of deformation and microstructure.
The method produces a latent space that compactly and extensibly encodes the microstructure-response map.
This latent space and the semantic directions implicit in it are applicable to a variety of applications.
We demonstrated its ability to replicate DNS QoIs, push forward microstructure uncertainty, and simulate functional gradation of stochastic media.
The demonstrations employed a common AM material, AlSi$_{10}$Mg; however the proposed method is amenable to other microstructures such as those with multiple phases and/or with pores (i.e. a null material).
In materials such as those generated by AM, defects can strongly affect the local stress gradients, and therefore scale separation.

The expediency of mapping structures to properities allowed us to interpret the maps and focus on the manipulation of the latent spaces to practical applications.
Encoding the structure-property map had straightforward implementation in a standard finite element code and was computationally efficient because the imprinting of latents on a mesh is done prior to simulation.
More general applications will likely require a more expressive structure-to-response map.
In future work we would like to extend the application of this method to microstructure ranging from random media with statistical characterization to highly structured lattice/metamaterials \cite{wang2020concurrent,del2020microstructure,hazeli2019microstructure} and also to more expressive latent distributions.
The method also has applications beyond forward propagation of material variability/uncertainty \cite{rizzi2019bayesian}, it could be indispensable in topology optimization \cite{xia2014concurrent} and material and structural optimization \cite{fernandez2022material,weeger2023inelastic}.

\section*{Acknowledgments}
The authors would like to thank Sara Dickens and Paul Specht, Sandia National Laboratories, for the microphotograph of AlSi$_{10}$Mg.
The simulations were performed with the SIERRA finite element simulator developed by Sandia National Laboratories.
Sandia National Laboratories is a multimission laboratory managed and operated by National Technology and Engineering Solutions of Sandia, LLC., a wholly owned subsidiary of Honeywell International, Inc., for the U.S. Department of Energy's National Nuclear Security Administration under contract DE-NA-0003525. This paper describes objective technical results and analysis. Any subjective views or opinions that might be expressed in the paper do not necessarily represent the views of the U.S. Department of Energy or the United States Government.



\end{document}